\documentclass[11pt,a4paper,twocolumn]{article}

\usepackage[utf8]{inputenc}
\usepackage[T1]{fontenc}
\usepackage{lmodern}
\usepackage{graphicx}
\usepackage{amsmath, amssymb}
\usepackage{booktabs}
\usepackage{multirow}
\usepackage{siunitx} 
\usepackage[hidelinks]{hyperref}
\usepackage{authblk}
\usepackage[
  a4paper,
  margin=0.75in,
  columnsep=0.25in  
]{geometry}
\usepackage{float} 
\usepackage{placeins}

\usepackage{enumitem}
\setlist{nosep, leftmargin=*, itemsep=2pt, parsep=0pt, topsep=4pt}

\usepackage{titlesec}
\titlespacing*{\section}{0pt}{10pt plus 2pt minus 2pt}{6pt plus 2pt minus 2pt}
\titlespacing*{\subsection}{0pt}{8pt plus 2pt minus 2pt}{4pt plus 2pt minus 2pt}
\titlespacing*{\subsubsection}{0pt}{6pt plus 2pt minus 2pt}{3pt plus 2pt minus 2pt}

\emergencystretch=\maxdimen
\title{When Do Domain-Specific Foundation Models Justify Their Cost? A Systematic Evaluation Across Retinal Imaging Tasks}

\author{
  D\'avid Isztl\textsuperscript{1,2},
  Tahm Spitznagel\textsuperscript{1,2},
  G\'abor M\'ark Somfai\textsuperscript{1,2},
  Rui Santos\textsuperscript{1,2,*}\\
  \textsuperscript{1}Stadtspital Z\"urich, Department of Ophthalmology, Z\"urich, Switzerland\\
  \textsuperscript{2}Spross Research Institute, Z\"urich, Switzerland\\
  *Corresponding author: rui.santos@stadtspital.ch
}

\date{\today}

\begin{document}

\twocolumn[
\begin{@twocolumnfalse}
\maketitle

\begin{abstract}
\noindent
Large vision foundation models have been widely adopted for retinal disease classification without systematic evidence justifying their parameter requirements. In the present work we address two critical questions: First, are large domain-specific foundation models essential, or do compact general-purpose architectures suffice? Second, does specialized retinal pretraining justify its computational cost? To answer this, we benchmark initialization strategies across four retinal imaging classification tasks spanning Optical Coherence Tomography (OCT) and Color Fundus Photography (CFP) modalities: 8-class OCT classification, 3-class diabetic macular edema (DME), 5-class diabetic retinopathy (DR), and 3-class glaucoma (GL) detection. We evaluate 12--13 model configurations per task, including vision transformers (22.8M--86.6M parameters), Swin Transformers (27.6M--28.3M), ConvNeXt (28.6M), and the domain-specific RETFound models (303M), under identical training conditions. Our results challenge prevailing assumptions: First, we demonstrate that pretraining provides universal benefits (5.18--18.41\% improvement), scaling with task difficulty. Second, compact architectures (27--29M) dominate Pareto frontiers; SwinV2-tiny achieves top-1 performance on three datasets. Third, RETFound (303M) justifies its computational cost only for challenging DR grading (accuracy of 71.15\%), while ImageNet pretraining proves to be sufficient with all other tasks (DME accuracy: 99.24\%, OCT accuracy: 97.96\%). CFP tasks show larger pretraining accuracy gains (9.13--18.41\%) than OCT (5.18\%). Thus, the evidence suggests that compact general-purpose models deliver near-optimal performance for most retinal classification tasks; specialized foundation models warranted only for fine-grained discrimination under extreme class imbalance.
\end{abstract}

\vspace{2em}
\end{@twocolumnfalse}
]

\section{Introduction}
\label{sec:intro}

Retinal imaging provides non-invasive access to vascular and neural tissue within the eye, enabling early detection and monitoring of sight-threatening diseases including diabetic retinopathy and macular edema, glaucoma. With over 500 million people affected by diabetes worldwide \cite{SUN2022109119} and glaucoma as the leading cause of irreversible blindness \cite{THAM20142081}, automated classification systems offer critical potential for scaling screening programs and reducing diagnostic burdens in resource-limited settings. Moreover, recent research in the field of oculomics amplifies the use of retinal imaging for non-invasive screening of systemic conditions such as cardiovascular and neurological diseases \cite{wagner2020insights,pattersonoculomics,zhu2025oculomics}.

Deep learning models have achieved expert-level performance on numerous retinal imaging tasks, with convolutional neural networks \cite{he2016deep} and vision transformers demonstrating remarkable accuracy across classification of diverse pathologies. The recent surge in vision foundation models has established a prevailing narrative: domain-specific pretraining on large-scale medical imaging datasets is essential, and larger models consistently outperform smaller architectures. Further research efforts have focused on developing specialized retinal foundation models such as RETFound with 303M parameters \cite{zhou2023foundation,zhou2025generalistversusspecialistvision}, under the assumption that visual features learned from generic natural images cannot transfer effectively to specialized medical imaging domains, necessitating domain-specific pretraining on large-scale retinal datasets for clinical-grade performance. This has driven substantial computational investments in self-supervised pretraining on either public or proprietary retinal imaging datasets.

However, systematic evidence supporting these assumptions is very limited \cite{taher2021systematicbenchmarkinganalysistransfer,bahr2024deep,huang2025systematic}. Prior studies have evaluated single architectures on single datasets, making it impossible to disentangle task-specific findings from generalizable insights about model scaling and specialization. Thus, critical questions have been left unanswered: Do compact architectures (27--29M parameters) suffice for most retinal imaging tasks, or are large foundation models essential? When does domain-specific retinal pretraining justify its computational cost compared to general-purpose ImageNet \cite{deng2009imagenet,russakovsky2015imagenet} initialization? How do these trade-offs vary across imaging modalities (Optical Coherence Tomography [OCT] vs.\ Color Fundus Photography [CFP]) and task difficulties, namely balanced multi-class vs.\ severely imbalanced ordinal grading?

We hypothesize that: compact general-purpose architectures (27--29M parameters) will achieve performance comparable to large domain-specific foundation models (300M+) across most retinal imaging tasks, challenging the prevailing ``bigger is better'' paradigm.

This study provides the first comprehensive evaluation of backbone initialization strategies and architectures spanning two modalities (OCT and CFP) across distinct retinal imaging tasks, of which, DR, DME, and Glaucoma for CFP. While AMD, CNV, CSR, DME, DR, DRUSEN and MH for OCT. Featuring varying complexity levels (balanced multi-class, few-class, ordinal with severe imbalance, and 3-class). We systematically evaluate 12--13 model configurations per dataset, including supervised and self-supervised vision transformers (22.8M--86.6M parameters), hierarchical Swin Transformers (27.6--28.3M), modern convolutional architectures (ConvNeXt, 28.6M), and domain-specific foundation models (RETFound, 303M). All models were trained under identical conditions using a unified training platform, isolating the effects of architecture and initialization from confounding training procedure variations.

Our key contributions are:
\begin{itemize}
  \item \textbf{Challenging scaling assumptions}: Pareto frontier analysis demonstrating that compact models (22.8--29M parameters) dominate accuracy-size trade-offs across all tasks, contradicting the prevailing "bigger is better" paradigm in foundation model development for ophthalmology.

  \item \textbf{Quantifying domain-specific pretraining value}: Large-scale retinal foundation models such as RETFound (303M), provide advantages only for the most challenging ordinal grading task (DR).

  \item \textbf{Multi-task generalization analysis}: Systematic comparison of pretraining benefits across multiple eye diseases (AMD, CNV, CSR, DME, DR, DRUSEN, MH, GL) spanning two imaging modalities and four difficulty levels, revealing when specialized models are warranted versus when compact general-purpose architectures suffice.

  \item \textbf{Architecture-modality interactions}: Evidence that ImageNet features transfer more effectively to color fundus photography (9.13--18.41\% pretraining benefit) than OCT imaging (5.18\%), thus informing resource allocation strategies across imaging modalities.
\end{itemize}

Our findings challenge current academic assumptions for retinal imaging AI model development. Compact hierarchical architectures such as SwinV2-tiny (27.6M) achieve top-2 performance on all four tasks, while domain-specific foundation models justify their computational investment only for fine-grained severity grading under extreme class imbalance. These results suggest that the drive towards ever-larger specialized foundation models may be premature, perhaps even unnecessary, with compact general-purpose architectures delivering near-optimal performance for most ophthalmological applications.

\newpage

\section{Methods}
\label{sec:methods}

\subsection{Datasets}
We evaluated models on four retinal imaging datasets: AMD, CNV, CSR, DME, DR, DRUSEN, MH (Retinal OCT C8, 24,000 images, 8 balanced disease categories), DR (EAM unified dataset, 143,669 images, 5-class ordinal ICDR scale with significant imbalance), DME (IDRiD + Messidor-2, 2,264 images, 3-class severity grading), and GL (AIROGS + PAPILA, 114,381 images, 3-class glaucoma detection). See Appendix~\ref{app:dataset_details} for complete dataset descriptions, and Table~\ref{tab:datasets} for dataset characteristics.

\subsection{Experimental Design}
We conducted a systematic comparison of backbone initialization strategies and architectures to evaluate their impact on retinal disease classification performance across all four datasets.

\subsubsection{Backbone Architectures}
Six backbone architectures were evaluated, spanning different design paradigms (Table~\ref{tab:backbones}). The selection includes supervised and self-supervised Vision Transformers, hierarchical Swin Transformers, modern convolutional architectures (ConvNeXt), and domain-specific models pretrained on retinal imaging data (RETFound). All pretrained models were obtained from the HuggingFace model hub; exact repository identifiers for reproducibility are provided in Table~\ref{tab:backbones_hf} (Appendix).

\begin{table*}[t]
\centering
\caption{Backbone architectures evaluated in this study. All models evaluated with both pretrained and from-scratch initialization, except RETFound and DINOv2 variants which were evaluated with pretrained weights only due to the computational demands of their self-supervised pretraining procedures.}
\label{tab:backbones}
\small
\begin{tabular}{@{}llcl@{}}
\toprule
\textbf{Architecture Family} & \textbf{Model} & \textbf{Parameters} \\
\midrule
Vision Transformer \cite{dosovitskiy2020image} & ViT-base & 86.6M \\
\midrule
\multirow{2}{*}{Self-supervised ViT \cite{oquab2024dinov2learningrobustvisual,darcet2023vision}} & DinoV2-small & 22.8M \\
 & DinoV2-small-reg & 86.6M \\
\midrule
\multirow{2}{*}{Swin Transformer \cite{liu2021swin,liu2022swin}} & Swin-tiny & 28.3M \\
 & SwinV2-tiny & 27.6M \\
\midrule
ConvNeXt \cite{liu2022convnet} & ConvNeXtV2-tiny & 28.6M \\
\midrule
\multirow{3}{*}{Domain-specific (RETFound)} & RETFound-MAE-OCT \cite{he2022masked,zhou2023foundation,zhou2025generalistversusspecialistvision} & 303M \\
 & RETFound-DinoV2-CFP \cite{zhou2023foundation,zhou2025generalistversusspecialistvision} & 303M \\
 & RETFound-MAE-CFP \cite{he2022masked,zhou2023foundation,zhou2025generalistversusspecialistvision} & 303M \\
\bottomrule
\end{tabular}
\end{table*}

\subsubsection{Initialization Strategies}
Two initialization approaches were compared: (1) \textbf{Pretrained}, where backbones were initialized with ImageNet-1k pretrained weights (or domain-specific retinal pretraining for RETFound models), with all layers fine-tuned end-to-end; (2) \textbf{From scratch}, where backbones were trained from random initialization. Four architectures (ViT-base, Swin-tiny, SwinV2-tiny, ConvNeXtV2-tiny) were evaluated with both strategies, while DINOv2 variants and RETFound models used pretrained weights only. This yielded 13 configurations for OCT and 12 for each CFP task (DME, DR, GL).

\subsubsection{Training Procedures}
All models were trained using a unified platform ensuring direct comparison. We employed AdamW optimization \cite{loshchilov2017decoupled} with cosine learning rate schedules and warmup (10\% of training), model-specific learning rates (2e-4 to 1e-3, tuned per architecture), weight decay 0.05, and gradient clipping (max norm 1.0). Effective batch size was 256 across all experiments via gradient accumulation. Training ran for 100 epochs without early stopping, with final results reporting best validation accuracy. Data augmentation was minimal (resize, center crop, normalization) to isolate architecture and initialization effects. Cross-entropy loss was used without class weighting. Training employed mixed-precision (bfloat16), deterministic seeds (42), and NVIDIA GPUs. See Appendix~\ref{app:training_details} for complete training infrastructure specifications.

\subsection{Evaluation Metrics}
We assessed model performance using five complementary metrics: (1) \textbf{Accuracy}, the top-1 classification accuracy; (2) \textbf{AUROC (macro)}, the macro-averaged Area Under the ROC Curve \cite{hanley1982meaning,fawcett2006introduction}, a threshold-independent discrimination measure; (3) \textbf{F1-score (macro)}, the macro-averaged harmonic mean of precision and recall \cite{powers2011evaluation}; (4) \textbf{Cohen's Kappa}, an agreement measure adjusted for chance \cite{cohen1960coefficient}, accounting for class imbalance; (5) \textbf{Mean Average Precision (mAP)}, which summarizes precision-recall curves. See Appendix~\ref{app:metrics_details} for complete metric definitions.

\subsection{Statistical Analysis}
Pretrained vs.\ scratch differences were assessed using independent Mann-Whitney U tests ($\alpha = 0.05$) \cite{mann1947test}. Models were ranked on accuracy, AUROC, and F1-score, with average rank across metrics providing an overall performance indicator. Parameter efficiency was defined as accuracy per 100M parameters. Pareto frontier analysis \cite{miettinen1999nonlinear} identified models that are optimal in the accuracy-size trade-off space (no other model achieves higher accuracy without more parameters). To quantify linear relationships between model size and performance, Pearson correlation coefficients \cite{pearson1895notesonregressionand} were computed for parameter count versus each performance metric (accuracy, AUROC, F1-score). Pearson correlation measures the strength and direction of linear association between two variables, ranging from -1 (perfect negative linear relationship) through 0 (no linear correlation) to +1 (perfect positive linear relationship). Statistical significance was assessed at $\alpha = 0.05$. See Appendix~\ref{app:statistical_details} for detailed statistical methodology.

\newpage

\section{Results}
\label{sec:results}

We present comprehensive results across four distinct eye disease classification tasks: Optical Coherence Tomography (OCT) imaging, and three Color Fundus Photography (CFP) tasks targeting Diabetic Macular Edema (DME), Diabetic Retinopathy (DR), and Glaucoma (GL). For each task, we evaluate the impact of backbone architecture, pretraining strategy, and model size on classification performance.

\subsection{Classification Results from OCT data}
\label{sec:results:oct}

\subsubsection{Overall Performance}
\begin{figure*}[!ht]
\centering
\includegraphics[width=0.85\textwidth]{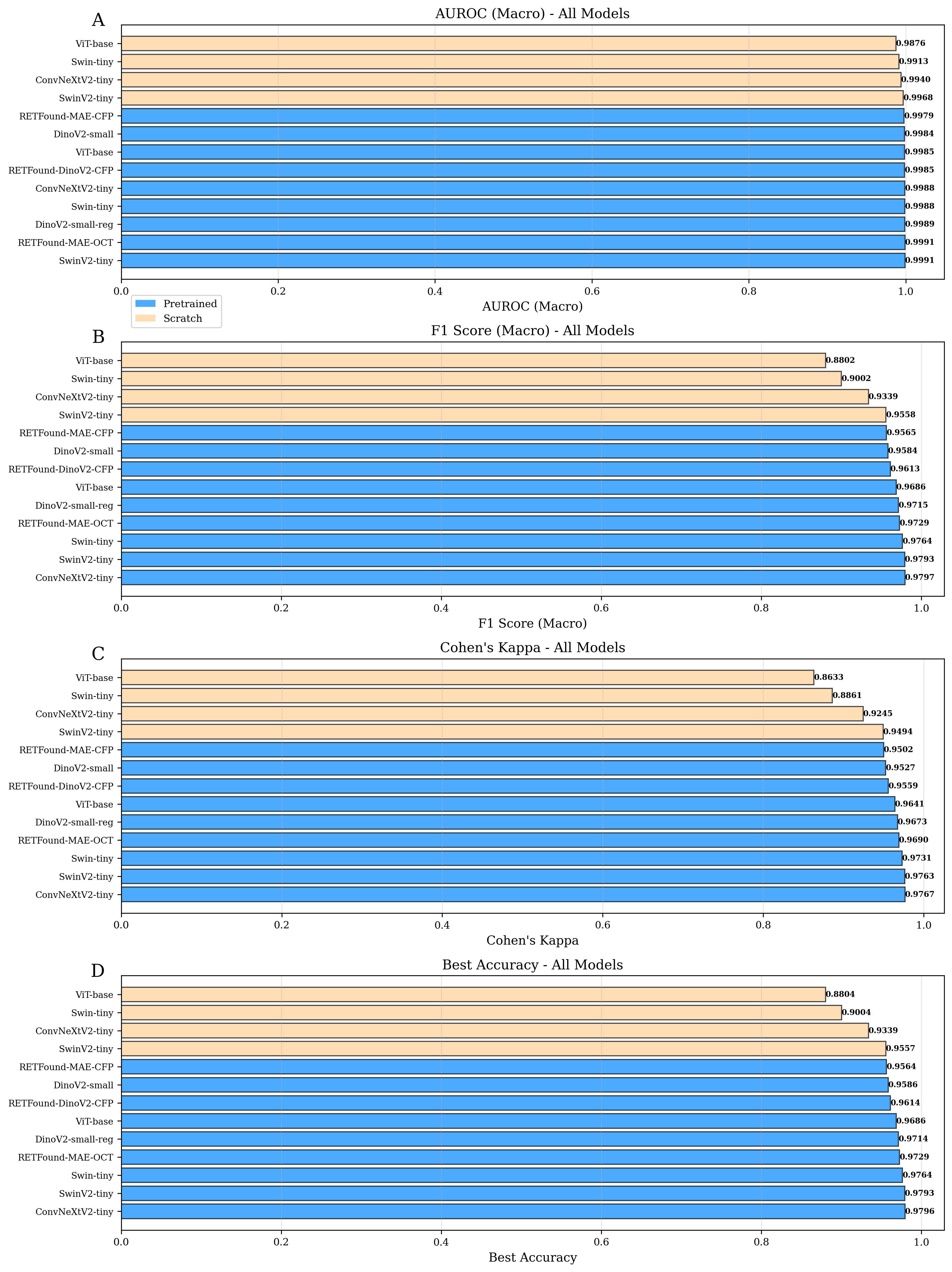}
\caption{OCT: Multi-panel horizontal bar chart comparing all 13 model configurations across four key metrics. Each panel shows models on the Y-axis (sorted by performance, bottom: best) with metric values [0.0--1.0] on the X-axis. Color coding distinguishes pretrained (blue bars) from scratch-trained (green bars) models. \textbf{(A) AUROC (macro)}: Threshold-independent discrimination; pretrained models cluster at 0.9979--0.9991, while scratch models range 0.9876--0.9968. All models achieve excellent AUROC >0.985, indicating strong class separability for the 8-class OCT. \textbf{(B) F1-Score (macro)}: Harmonic mean of precision/recall; pretrained models 0.9565--0.9797, scratch 0.8802--0.9558. F1-Score shows the same pretrained-scratch separation pattern as accuracy. \textbf{(C) Cohen's Kappa}: Chance-adjusted agreement; pretrained 0.9502--0.9767, scratch 0.8633--0.9494. The 8--10 percentage point gap demonstrates pretraining's value beyond chance-level agreement. \textbf{(D) Accuracy}: Pretrained models dominate top positions (0.9565--0.9796), with ConvNeXtV2-tiny and SwinV2-tiny leading. All pretrained models exceed 0.95, while scratch models span 0.8804--0.9557. Consistent stratification across all panels confirms that ImageNet initialization provides universal benefits regardless of metric choice.}
\label{fig:validation_metrics_oct}
\end{figure*}

We evaluated 13 model configurations on OCT classification, with model sizes ranging from 22.8M to 303M parameters. The highest accuracy was achieved by the pretrained ConvNeXtV2-tiny (28.6M parameters) at 97.96\% validation accuracy, 99.88\% AUROC, and 97.97\% F1-score. Table~\ref{tab:top_models} summarizes the top-performing models. Figure~\ref{fig:validation_metrics_oct} provides a comprehensive comparison of validation metrics across all 13 models.

\begin{table*}[t]
\centering
\caption{Top-5 performing models on OCT classification. All top models use pretrained initialization. \textbf{Acc}: Top-1 validation accuracy. \textbf{AUROC}: Area Under the Receiver Operating Characteristic curve, macro-averaged across classes (threshold-independent performance measure). \textbf{F1}: F1-score, macro-averaged across classes (harmonic mean of precision and recall). \textbf{Avg. Rank}: Average rank across the three metrics (lower is better); models ranked separately on each metric, then averaged.}
\label{tab:top_models}
\small
\begin{tabular}{@{}lcccccc@{}}
\toprule
\textbf{Model} & \textbf{Pretrained} & \textbf{Params} & \textbf{Acc (\%)} & \textbf{AUROC} & \textbf{F1} & \textbf{Avg. Rank} \\
\midrule
SwinV2-tiny & Yes & 27.6M & 97.93 & 0.9991 & 0.9793 & 1.67 \\
ConvNeXtV2-tiny & Yes & 28.6M & 97.96 & 0.9988 & 0.9797 & 2.33 \\
Swin-tiny & Yes & 28.3M & 97.64 & 0.9988 & 0.9764 & 3.33 \\
RETFound-MAE-OCT & Yes & 303M & 97.29 & 0.9991 & 0.9729 & 3.33 \\
DinoV2-small-reg & Yes & 86.6M & 97.14 & 0.9989 & 0.9715 & 4.33 \\
\bottomrule
\end{tabular}
\end{table*}

\subsubsection{Impact of Pretraining}
\begin{figure*}[!ht]
\centering
\includegraphics[width=1.0\textwidth]{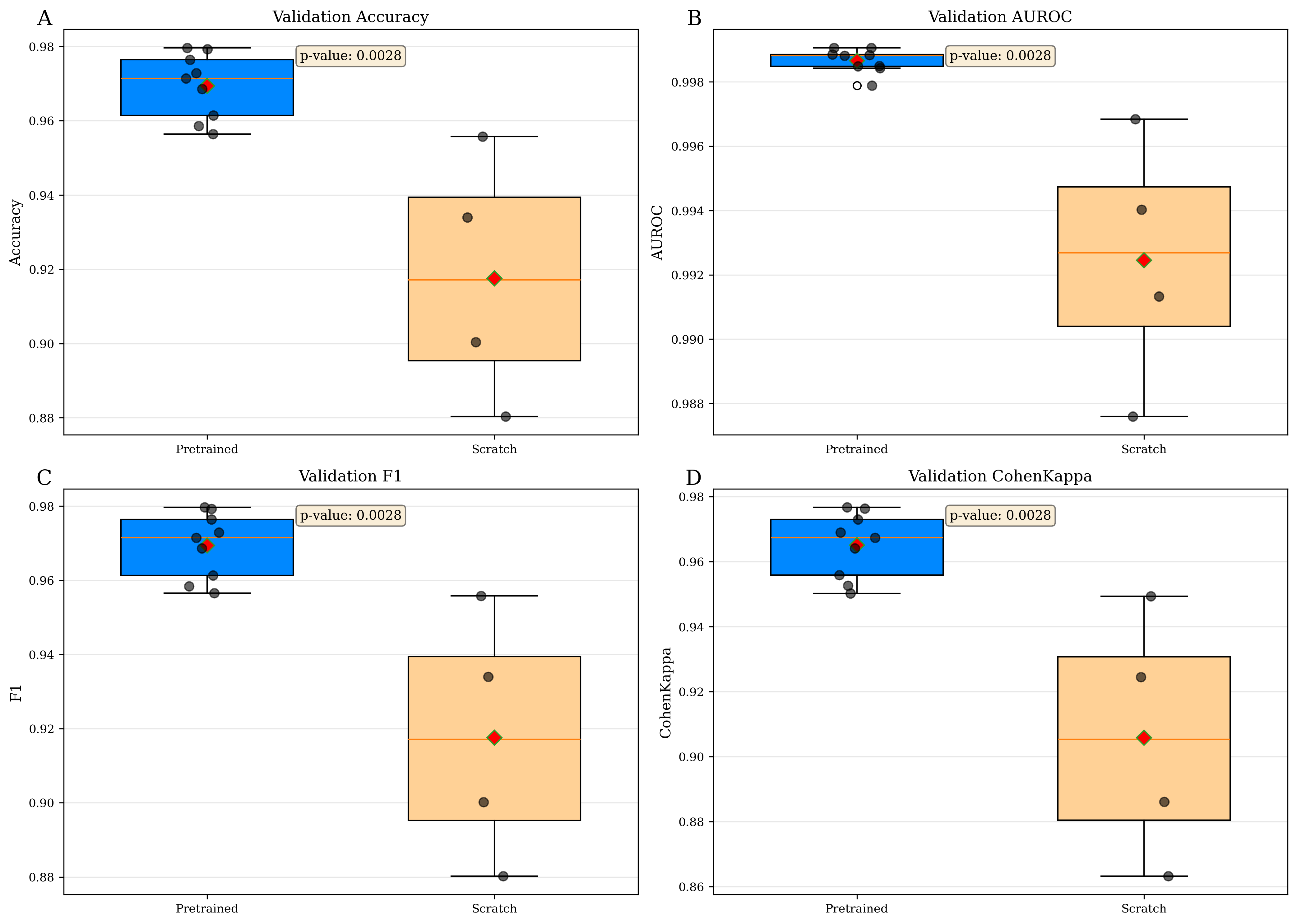}
\caption{OCT: Multi-panel box plot comparison of pretrained vs scratch-trained models across four key metrics. Each panel shows box-and-whisker plots where boxes represent the interquartile range (IQR, 25th--75th percentiles), horizontal lines show medians, whiskers extend to 1.5$\times$IQR, and individual points indicate outliers. \textbf{(A) Best Validation Accuracy}: Pretrained models (n=9) achieve mean 96.94\% $\pm$ 0.88\% vs scratch (n=4) 91.76\% $\pm$ 3.37\%, representing a 5.18 percentage point improvement (p < 0.05, Mann-Whitney U test) with 3.8$\times$ variance reduction. \textbf{(B) AUROC (macro)}: Pretrained models achieve 99.87\% $\pm$ 0.04\% vs scratch 99.24\% $\pm$ 0.39\%, a 0.63\% improvement (p < 0.05) demonstrating superior threshold-independent discrimination. \textbf{(C) F1-Score (macro)}: Pretrained 96.94\% $\pm$ 0.88\% vs scratch 91.76\% $\pm$ 3.38\%, mirroring accuracy patterns. \textbf{(D) Cohen's Kappa}: Pretrained 96.50\% $\pm$ 1.00\% vs scratch 90.58\% $\pm$ 3.85\%, showing 5.92 percentage point improvement in chance-adjusted agreement. Across all metrics, pretrained models show dramatically reduced variance and higher central tendency, indicating that ImageNet initialization provides universal, consistent benefits for the 8-class OCT classification.}
\label{fig:pretrained_vs_scratch_oct}
\end{figure*}

Pretrained models significantly outperformed those trained from scratch across all metrics (Table~\ref{tab:pretrained_comparison}). Pretrained models achieved mean best validation accuracy of $0.9694 \pm 0.0088$ (n=9) compared to $0.9176 \pm 0.0337$ (n=4) for scratch-trained models, representing an absolute improvement of 5.18 percentage points. This difference was statistically significant (independent Mann-Whitney U test, $p < 0.05$). Similar patterns were observed for AUROC (pretrained: $0.9987 \pm 0.0004$ vs.\ scratch: $0.9924 \pm 0.0039$) and F1-score (pretrained: $0.9694 \pm 0.0088$ vs.\ scratch: $0.9176 \pm 0.0338$), demonstrating the consistent value of transfer learning for OCT classification. Figure~\ref{fig:pretrained_vs_scratch_oct} illustrates these differences across all metrics.

\begin{table*}[t]
\centering
\caption{Comparison of pretrained vs.\ scratch-trained models across all metrics. Values reported as mean $\pm$ standard deviation. Statistical significance assessed using independent two-sample Mann-Whitney U test. \textbf{AUROC (Macro)}: Area under ROC curve, macro-averaged (equal weight per class, threshold-independent). \textbf{F1-Score (Macro)}: Harmonic mean of precision and recall, macro-averaged. \textbf{Cohen's Kappa}: Agreement metric adjusted for chance agreement, accounting for class imbalance ($\kappa = 1$ indicates perfect agreement, $\kappa = 0$ indicates chance-level agreement). \textbf{Best Validation Acc}: Highest accuracy achieved across all training epochs.}
\label{tab:pretrained_comparison}
\small
\begin{tabular}{@{}lcccc@{}}
\toprule
\textbf{Metric} & \textbf{Pretrained (n=9)} & \textbf{Scratch (n=4)} & \textbf{Difference} & \textbf{p-value} \\
\midrule
AUROC (Macro) & 0.9987 $\pm$ 0.0004 & 0.9924 $\pm$ 0.0039 & +0.63\% & 0.0028 \\
F1-Score (Macro) & 0.9694 $\pm$ 0.0088 & 0.9176 $\pm$ 0.0338 & +5.18\% & 0.0028 \\
Cohen's Kappa & 0.9650 $\pm$ 0.0100 & 0.9058 $\pm$ 0.0385 & +5.92\% & 0.0028 \\
Best Validation Acc & 0.9694 $\pm$ 0.0088 & 0.9176 $\pm$ 0.0337 & +5.18\% & 0.0028 \\
\bottomrule
\end{tabular}
\end{table*}

\subsubsection{Model Rankings}
\begin{figure*}[!ht]
\centering
\includegraphics[width=1.0\textwidth]{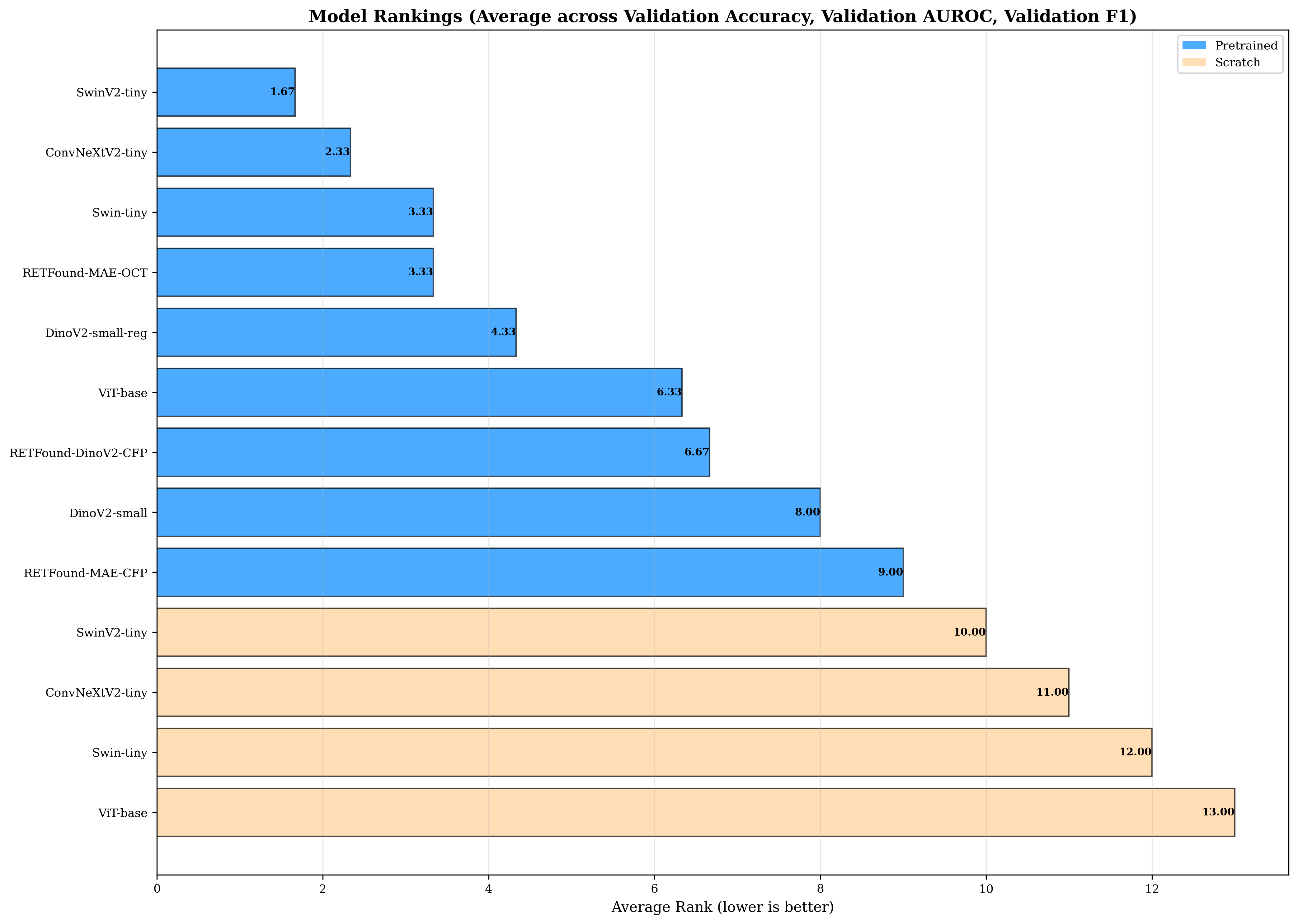}
\caption{OCT: Aggregate model rankings based on average rank across three key metrics (accuracy, AUROC, F1-score). Each model was ranked separately on each metric (rank 1 = best), then average rank computed to identify overall top performers while reducing single-metric bias. Lower average rank indicates superior overall performance. The visualization reveals a clear stratification: all nine pretrained models occupy ranks 1--9, while all four scratch-trained models occupy ranks 10--13, with zero overlap. SwinV2-tiny achieves the best average rank (1.67), followed by ConvNeXtV2-tiny (2.33), both compact hierarchical architectures in the 27--29M parameter range. The 303M RETFound-MAE-OCT model ranks only 4th (average rank 3.33), demonstrating that domain-specific pretraining on retinal images provides no advantage over ImageNet initialization for this OCT classification task.}
\label{fig:rankings_oct}
\end{figure*}

Models were ranked by performance across three key metrics (accuracy, AUROC, F1-score), as shown in Figure~\ref{fig:rankings_oct}. The Swin-tiny (pretrained) achieved the best average rank (1.67), followed closely by ConvNeXtV2-tiny (pretrained, rank 2.33). Notably, all top-5 ranked models were pretrained, with pretrained models occupying ranks 1--9 before any scratch-trained model appeared (rank 10).

\subsubsection{Parameter Efficiency}
\begin{figure*}[!ht]
\centering
\includegraphics[width=1.0\textwidth]{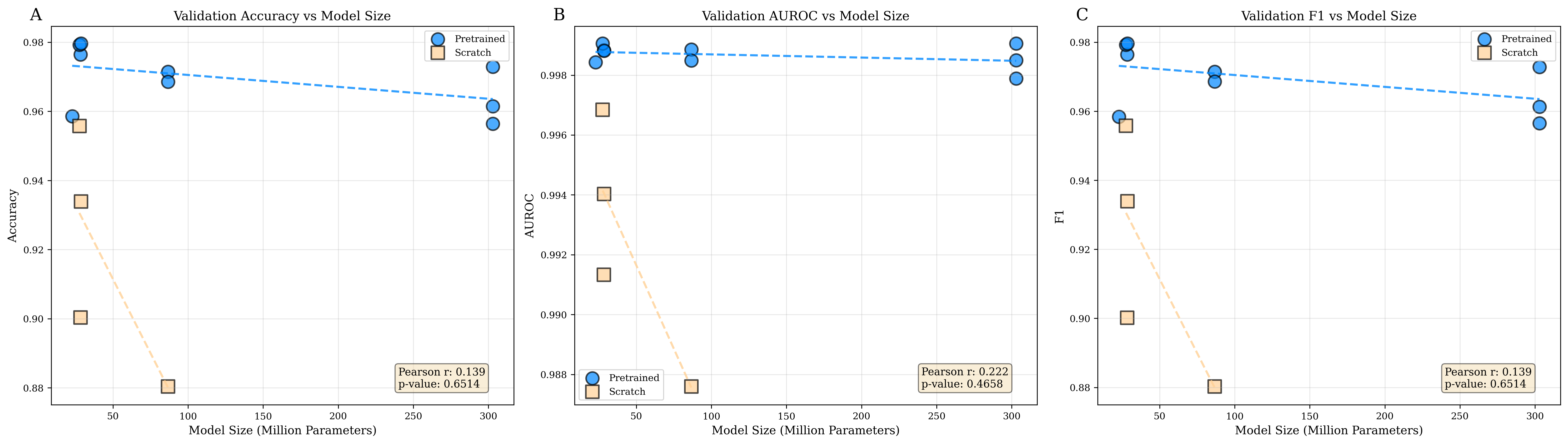}
\caption{OCT: Multi-panel scatter plots showing model performance versus parameter count across three metrics. Pretrained models shown in blue, scratch-trained in green. Each point represents one model configuration. Pearson correlation statistics (r, p-value) quantify the linear relationship between model size and performance. \textbf{(A) Accuracy vs Parameter Count}: Pretrained models (0.9564--0.9796) consistently outperform scratch models (0.8804--0.9557) by 5--9 percentage points regardless of size. Performance saturates at 28--30M parameters (~0.98 accuracy); larger models (86.6M ViT, 303M RETFound) provide minimal gains, demonstrating diminishing returns. The vertical separation between green/blue clusters visualizes the 5.18\% mean pretraining advantage. No significant correlation between size and accuracy ($p = 0.65$) confirms that larger models do not systematically outperform smaller ones. \textbf{(B) AUROC vs Parameter Count}: Similar saturation pattern; pretrained models cluster at 0.9979--0.9991, while scratch models range 0.9876--0.9968. No significant correlation ($p = 0.47$). \textbf{(C) F1-Score vs Parameter Count}: Mirrors accuracy patterns; pretrained 0.9565--0.9797, scratch 0.8633--0.9494. No significant correlation ($p = 0.65$). Consistent across panels: compact pretrained models (23--29M) achieve near-optimal performance, while scaling to 300M+ parameters yields negligible benefits.}
\label{fig:performance_size_oct}
\end{figure*}

\begin{figure*}[!ht]
\centering
\includegraphics[width=0.85\textwidth]{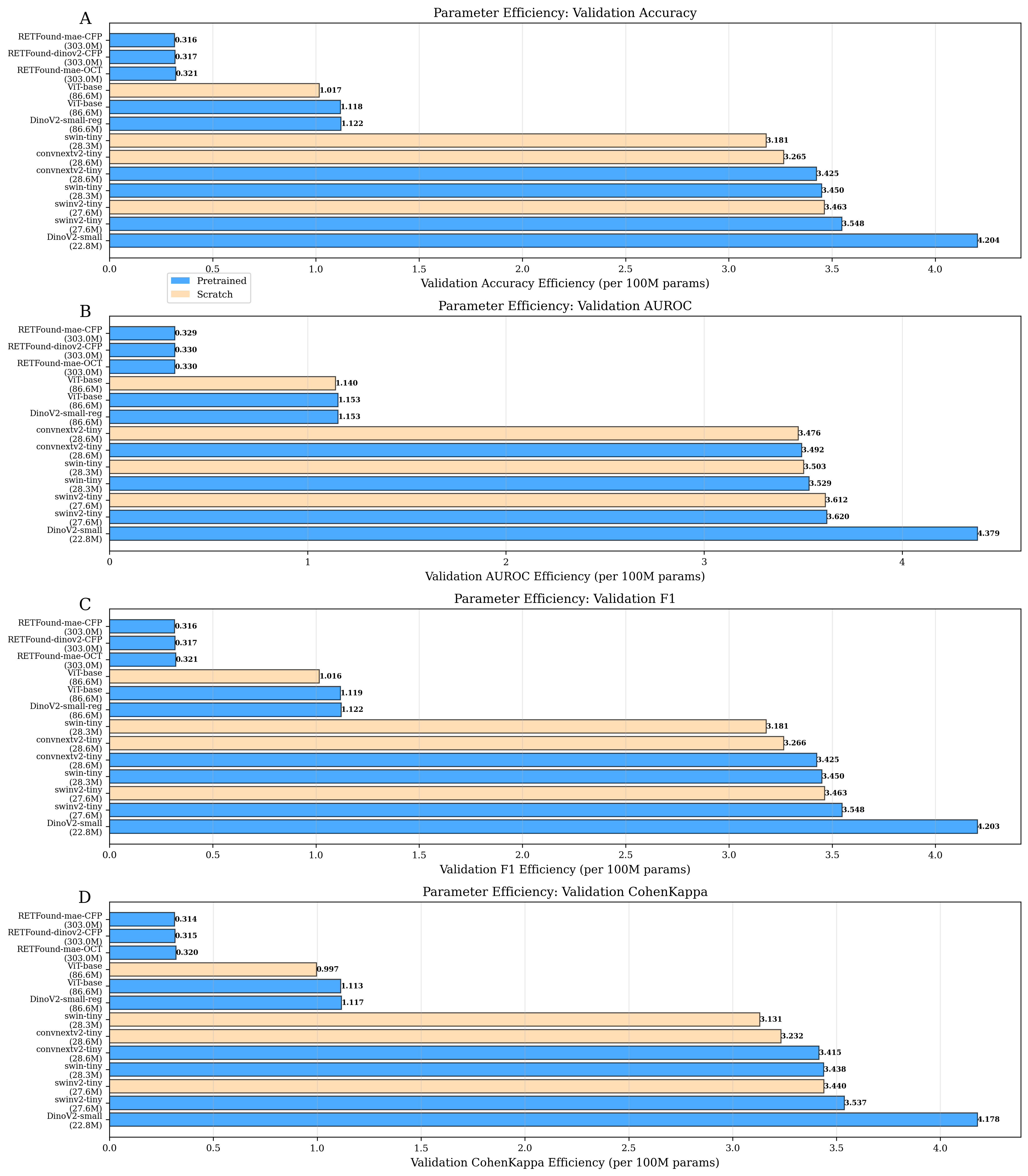}
\caption{OCT: Multi-panel horizontal bar chart showing parameter efficiency (metric value per 100M parameters) across all 13 model configurations. Each panel shows models on the Y-axis (with parameter counts) sorted by efficiency (bottom: most efficient), with efficiency values [0.0--5.0] on the X-axis. Color coding distinguishes pretrained (blue) from scratch (green) models. \textbf{(A) Accuracy Efficiency}: DinoV2-small leads with 4.204 points/100M params (95.86\% accuracy, 22.8M params)—only 2.1 percentage points below best while using 13$\times$ fewer parameters than RETFound. SwinV2-tiny (3.548, 97.93\%) and ConvNeXtV2-tiny (3.425, 97.96\%) provide optimal efficiency-performance balance. Scratch models show poorer efficiency. \textbf{(B) AUROC Efficiency}: Similar pattern with pretrained models dominating. \textbf{(C) F1 Efficiency}: DinoV2-small again leads (4.203), followed by compact pretrained models (3.4--3.6). \textbf{(D) Kappa Efficiency}: Pretrained models <4.2, scratch <3.4. Consistent across panels: compact pretrained models deliver superior efficiency; large specialized models (RETFound: 0.31--0.32 efficiency) show poor parameter utilization for this OCT classification task.}
\label{fig:efficiency_oct}
\end{figure*}

Parameter efficiency analysis revealed that smaller pretrained models delivered competitive performance relative to their size (Figure~\ref{fig:efficiency_oct}). The DinoV2-small (22.8M parameters) achieved the highest efficiency metric of 4.20 accuracy points per 100M parameters, with 95.86\% validation accuracy. Despite having 13$\times$ fewer parameters than the largest models (303M), it achieved 97.9\% of the best model's performance. Among models in the 27--29M parameter range, ConvNeXtV2-tiny and Swin-tiny demonstrated the best balance of size and accuracy, as illustrated in Figure~\ref{fig:performance_size_oct}. Pearson correlation analysis revealed no significant linear relationship between model size and performance metrics (accuracy: $p = 0.65$; AUROC: $p = 0.47$; F1-score: $p = 0.65$), demonstrating that larger models do not systematically outperform compact architectures on this OCT classification task.

\subsubsection{Pareto Frontier Analysis}
\begin{figure*}[!ht]
\centering
\includegraphics[width=1.0\textwidth]{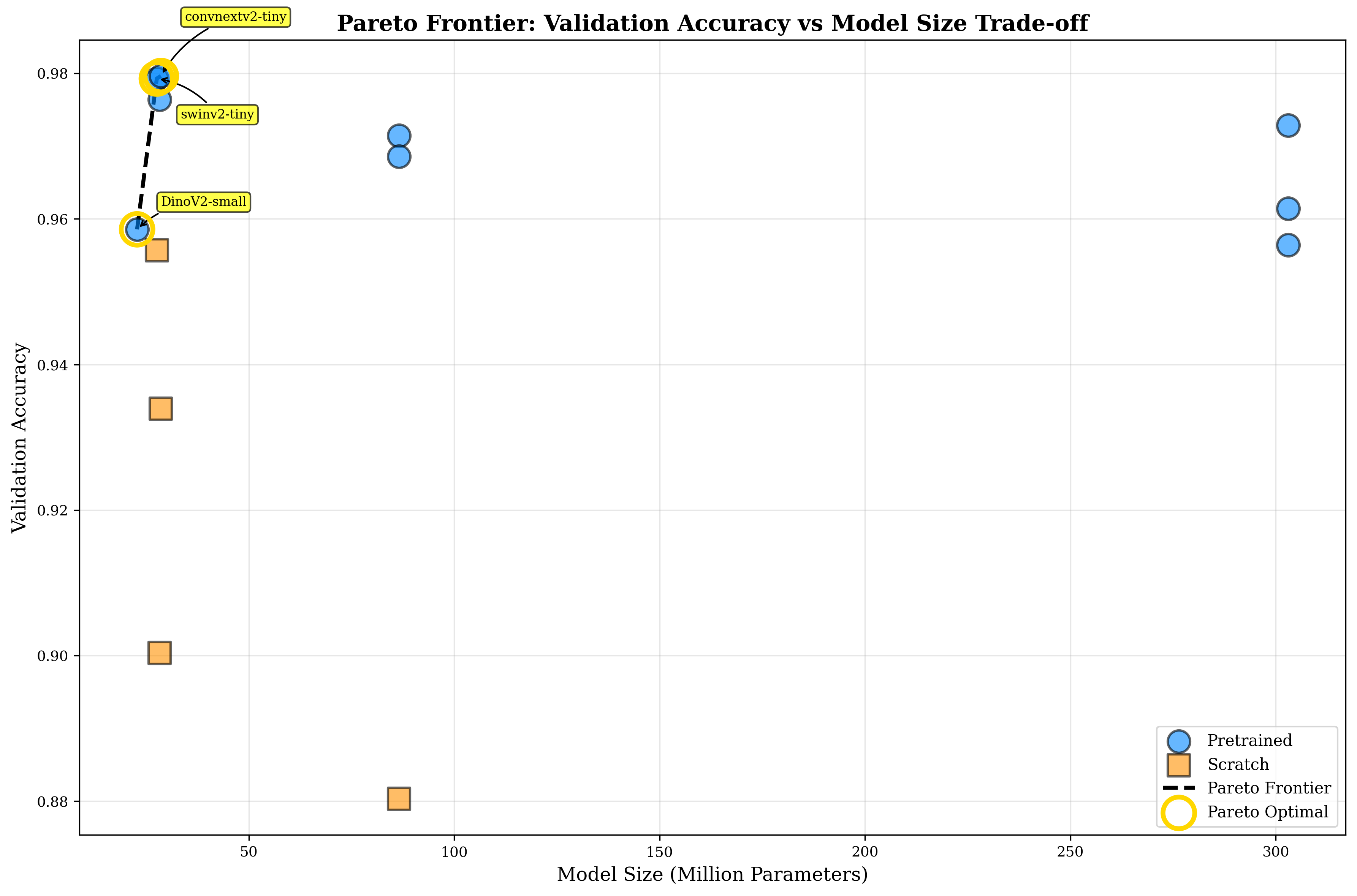}
\caption{OCT: Pareto frontier analysis identifying objectively optimal models in the accuracy-parameter trade-off space. Models on the frontier are Pareto-optimal: no other model achieves higher accuracy without requiring more parameters. The x-axis shows model size, while y-axis shows validation accuracy. Three models dominate the frontier: DinoV2-small (22.8M, 95.86\%), SwinV2-tiny (27.6M, 97.93\%), and ConvNeXtV2-tiny (28.6M, 97.96\%). Critically, all frontier models cluster in the 23--29M parameter range; larger models including RETFound-MAE-OCT (303M, 97.29\%) and DinoV2-small-reg (86.6M, 97.14\%) fall below the frontier, demonstrating that increased model size does not yield proportional accuracy gains for this 8-class OCT classification task. This challenges the assumption that 300M+ parameter foundation models are necessary for optimal performance.}
\label{fig:pareto_oct}
\end{figure*}

Pareto frontier analysis identified three objectively optimal models for the accuracy-size trade-off (Figure~\ref{fig:pareto_oct}):
\begin{itemize}
  \item \textbf{Small}: DinoV2-small (22.8M, 95.86\% accuracy), best for resource-constrained deployment
  \item \textbf{Medium}: Swin-tiny (27.6M, 97.93\% accuracy), optimal balance
  \item \textbf{Large}: ConvNeXtV2-tiny (28.6M, 97.96\% accuracy), maximum performance in compact form
\end{itemize}
Notably, the Pareto frontier was dominated by models in the 23--29M parameter range; with larger models (86.6M, 303M) not providing sufficient accuracy gains to justify their increased computational cost.

\subsubsection{Architecture-Specific Observations}
Among architectural families, Swin Transformers and ConvNeXt variants demonstrated the most consistent high performance when pretrained. The domain-specific RETFound models (303M parameters, pretrained on retinal images) did not outperform smaller ImageNet-pretrained models, suggesting that general-purpose visual representations transfer effectively to OCT classification. Vision Transformers showed the largest performance gap between pretrained and scratch configurations (ViT-base: 96.86\% pretrained vs.\ 88.04\% scratch), indicating strong dependence on initialization quality. See Appendix~\ref{app:oct_figures} for additional visualizations.

\subsection{Diabetic Macular Edema (DME) Classification Results from CFP data}
\label{sec:results:cfp_dme}

\subsubsection{Overall Performance}
\begin{figure*}[!ht]
\centering
\includegraphics[width=0.85\textwidth]{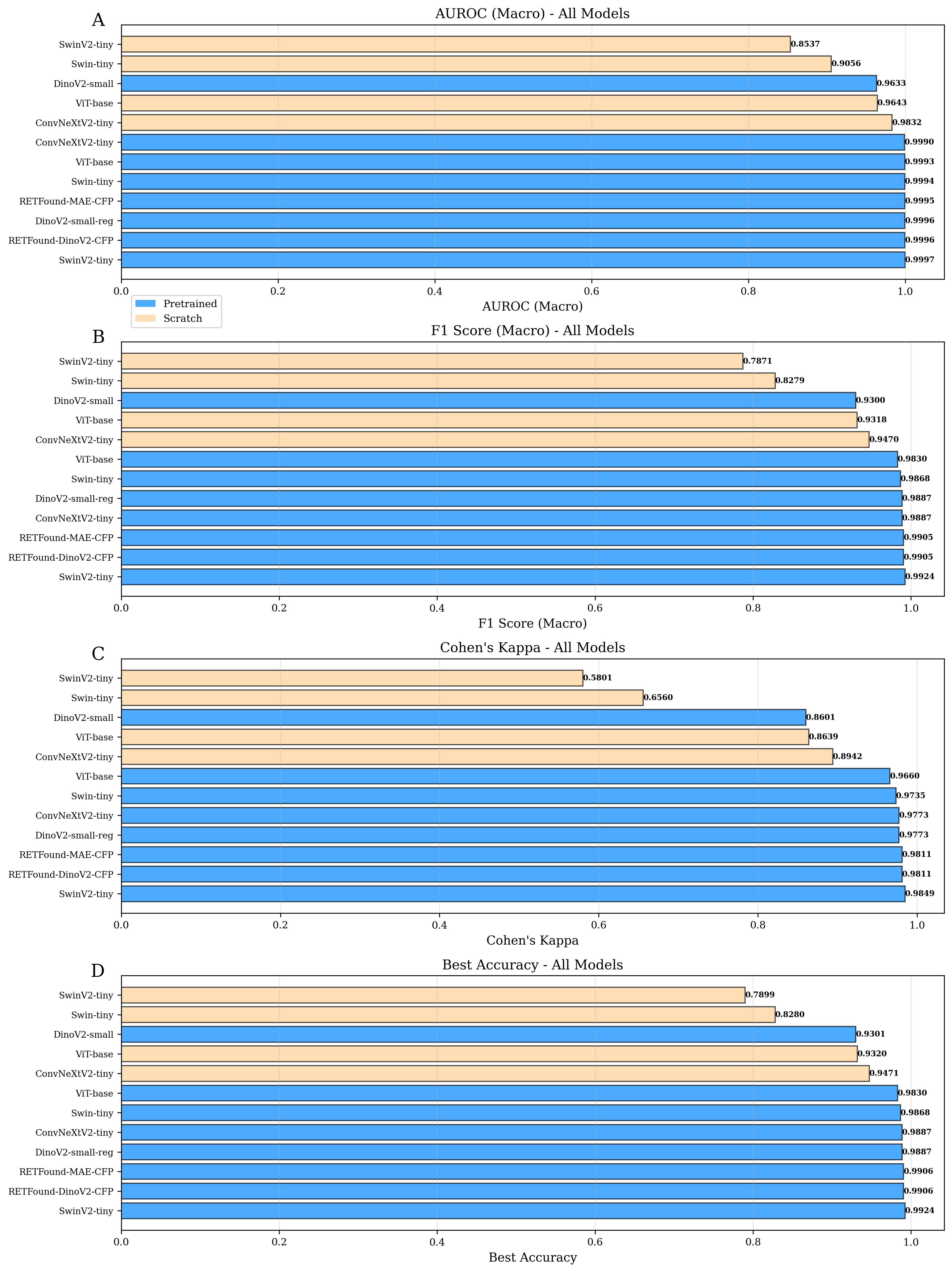}
\caption{DME: Multi-panel horizontal bar chart comparing all 12 model configurations for diabetic macular edema classification. Models on Y-axis (sorted by performance, bottom: best), metric values [0.0--1.0] on X-axis. \textbf{(A) AUROC (macro)}: Exceptional performance across all pretrained models (0.9633--0.9997), with SwinV2-tiny achieving 0.9997, near-perfect class discrimination. Scratch models show wider range (0.8537--0.9832). This is the highest AUROC task in the study. \textbf{(B) F1-Score (macro)}: SwinV2-tiny leads at 0.9924, with pretrained models clustering 0.9300--0.9924. Scratch models span 0.7871--0.9470, visualizing the 10.78\% pretraining benefit. \textbf{(C) Cohen's Kappa}: Substantial separation; pretrained 0.8601--0.9849, scratch 0.5801--0.8942. The 21.41\% mean improvement demonstrates critical value for chance-adjusted agreement on limited-data tasks. \textbf{(D) Accuracy}: SwinV2-tiny achieves 0.9924, outperforming 303M RETFound models (0.9906). All pretrained >0.93, scratch 0.7899--0.9471. Consistent stratification across all panels confirms general-purpose ImageNet pretraining suffices for this 3-class CFP task.}
\label{fig:validation_metrics_dme}
\end{figure*}

We evaluated 12 model configurations on DME classification (22.8M--303M parameters). The best-performing model was Swin-tiny (pretrained, 27.6M parameters), achieving 99.24\% validation accuracy, 99.97\% AUROC, and 99.24\% F1-score. Table~\ref{tab:top_models_dme} summarizes the top-performing models. Figure~\ref{fig:validation_metrics_dme} provides a comprehensive comparison of validation metrics across all 12 models.

\begin{table*}[t]
\centering
\caption{Top-5 performing models on DME classification. All top models use pretrained initialization. Metrics as defined in Table~\ref{tab:top_models}.}
\label{tab:top_models_dme}
\small
\begin{tabular}{@{}lcccccc@{}}
\toprule
\textbf{Model} & \textbf{Pretrained} & \textbf{Params} & \textbf{Acc (\%)} & \textbf{AUROC} & \textbf{F1} & \textbf{Avg. Rank} \\
\midrule
SwinV2-tiny & Yes & 27.6M & 99.24 & 0.9997 & 0.9924 & 1.00 \\
RETFound-DinoV2-CFP & Yes & 303M & 99.06 & 0.9996 & 0.9905 & 2.50 \\
RETFound-MAE-CFP & Yes & 303M & 99.06 & 0.9995 & 0.9905 & 3.00 \\
DinoV2-small-reg & Yes & 86.6M & 98.87 & 0.9996 & 0.9887 & 3.83 \\
ConvNeXtV2-tiny & Yes & 28.6M & 98.87 & 0.9990 & 0.9887 & 5.33 \\
\bottomrule
\end{tabular}
\end{table*}

\subsubsection{Impact of Pretraining}
\begin{figure*}[!ht]
\centering
\includegraphics[width=1.0\textwidth]{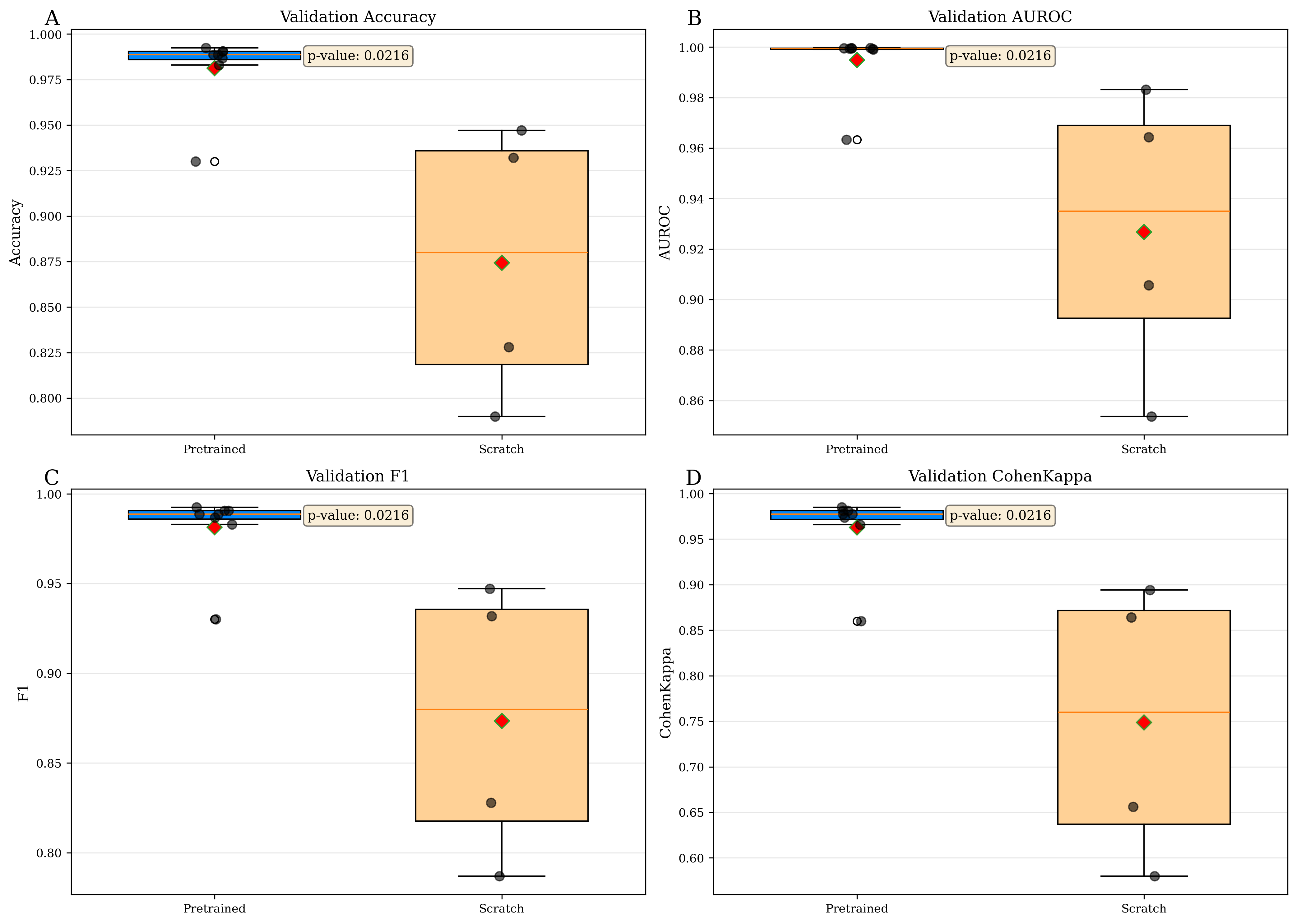}
\caption{DME: Multi-panel box plot comparison for 3-class diabetic macular edema severity grading. \textbf{(A) Best Validation Accuracy}: Pretrained models (n=8) achieve mean 98.13\% $\pm$ 2.09\% vs scratch (n=4) 87.43\% $\pm$ 7.72\%, representing a 10.70 percentage point improvement (p < 0.05), more than double the OCT advantage (5.18\%), indicating CFP-based tasks with limited data (2,264 images) benefit substantially more from ImageNet initialization. The substantial 3.7$\times$ variance reduction demonstrates critically improved training stability. \textbf{(B) AUROC (macro)}: Pretrained 99.49\% $\pm$ 1.28\% vs scratch 92.67\% $\pm$ 5.88\%, a 6.82 percentage point improvement (p < 0.05) showing substantially better class discrimination. \textbf{(C) F1-Score (macro)}: Pretrained 98.13\% $\pm$ 2.09\% vs scratch 87.35\% $\pm$ 7.82\%, mirroring the 10.78 percentage point accuracy gap. \textbf{(D) Cohen's Kappa}: Pretrained 96.27\% $\pm$ 4.18\% vs scratch 74.86\% $\pm$ 15.43\%, showing a substantial 21.41 percentage point improvement in chance-adjusted agreement, the largest Cohen's Kappa benefit across all tasks, reflecting the critical importance of pretraining for reliable performance on limited-data CFP classification.}
\label{fig:pretrained_vs_scratch_dme}
\end{figure*}

Pretrained models dramatically outperformed scratch-trained models on DME (Table~\ref{tab:pretrained_comparison_dme}, Figure~\ref{fig:pretrained_vs_scratch_dme}). Pretrained models achieved mean best validation accuracy of $0.9813 \pm 0.0209$ (n=8) versus $0.8743 \pm 0.0772$ (n=4) for scratch models, representing an absolute improvement of 10.70 percentage points (p < 0.05). This advantage substantially exceeds the 5.18\% gap observed for OCT, suggesting pretraining provides even greater benefit for CFP-based tasks, likely due to smaller dataset size and more effective ImageNet feature transfer to RGB imaging. Similar large gaps appeared in AUROC (pretrained: $0.9949 \pm 0.0128$ vs.\ scratch: $0.9267 \pm 0.0588$) and F1-score (pretrained: $0.9813 \pm 0.0209$ vs.\ scratch: $0.8735 \pm 0.0782$).

\begin{table*}[t]
\centering
\caption{Comparison of pretrained vs.\ scratch-trained models for DME. Metrics as defined in Table~\ref{tab:pretrained_comparison}.}
\label{tab:pretrained_comparison_dme}
\small
\begin{tabular}{@{}lcccc@{}}
\toprule
\textbf{Metric} & \textbf{Pretrained (n=8)} & \textbf{Scratch (n=4)} & \textbf{Difference} & \textbf{p-value} \\
\midrule
AUROC (Macro) & 0.9949 $\pm$ 0.0128 & 0.9267 $\pm$ 0.0588 & +6.82\% & 0.0216 \\
F1-Score (Macro) & 0.9813 $\pm$ 0.0209 & 0.8735 $\pm$ 0.0782 & +10.78\% & 0.0216 \\
Cohen's Kappa & 0.9627 $\pm$ 0.0418 & 0.7486 $\pm$ 0.1543 & +21.41\% & 0.0216 \\
Best Validation Acc & 0.9813 $\pm$ 0.0209 & 0.8743 $\pm$ 0.0772 & +10.70\% & 0.0216 \\
\bottomrule
\end{tabular}
\end{table*}

\subsubsection{Model Rankings and Efficiency}
\begin{figure*}[!ht]
\centering
\includegraphics[width=1.0\textwidth]{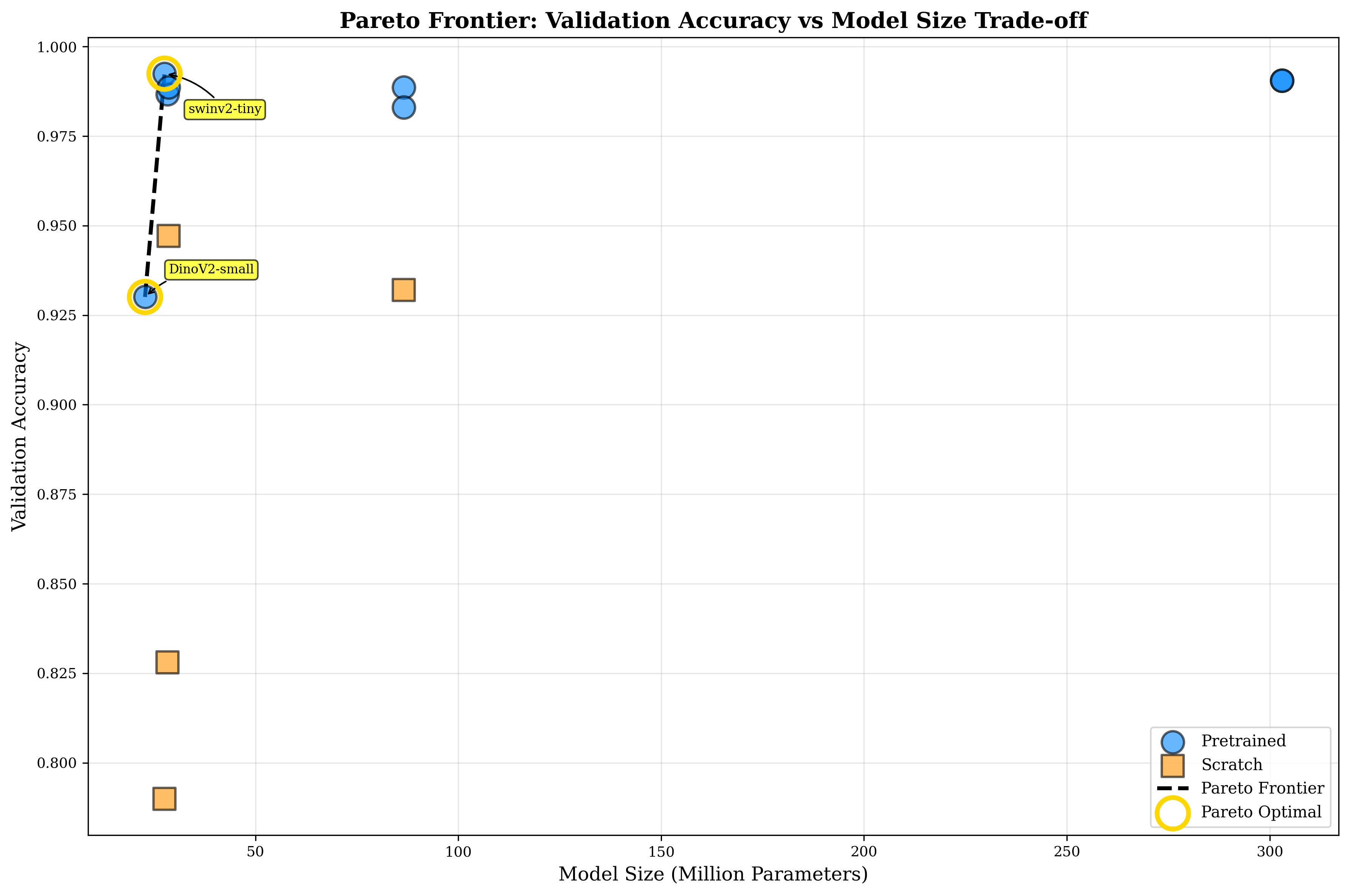}
\caption{DME: Pareto frontier analysis for diabetic macular edema classification. Only two models achieve Pareto optimality: DinoV2-small (22.8M, 93.01\% accuracy) and SwinV2-tiny (27.6M, 99.24\% accuracy). The frontier demonstrates an exceptionally steep accuracy gain between these points, 6.23 percentage points improvement for only 4.8M additional parameters, indicating that modest architecture scaling from 23M to 28M yields substantial performance benefits for this task. Beyond 28M parameters, no model provides accuracy gains justifying increased size; the 303M RETFound models achieve 99.06\% (0.18 points below SwinV2-tiny), falling well below the frontier. This result challenges the necessity of 300M+ parameter domain-specific foundation models for clinical grading tasks with clear diagnostic criteria.}
\label{fig:pareto_dme}
\end{figure*}

\begin{figure*}[!ht]
\centering
\includegraphics[width=1.0\textwidth]{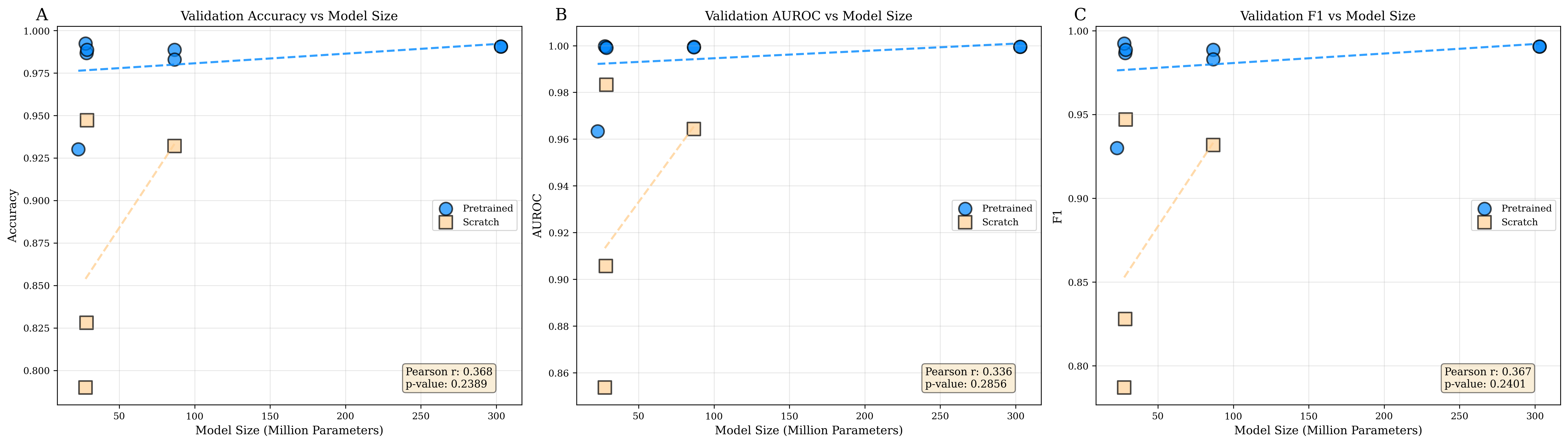}
\caption{DME: Multi-panel scatter plots showing model performance versus parameter count for diabetic macular edema classification. Pretrained (blue), scratch (green). Pearson correlation statistics (r, p-value) quantify the linear relationship between model size and performance. \textbf{(A) Accuracy vs Parameter Count}: Substantial 10.70 percentage point pretraining advantage; pretrained cluster tightly at 0.93--0.99, scratch spread 0.79--0.95. Performance saturates at 28M parameters (SwinV2-tiny, 0.9924); larger models (303M RETFound) provide negligible or negative gains. The best scratch model (ConvNeXtV2, 0.9471) still underperforms all pretrained models, demonstrating that limited data (2,264 images) benefits from pretraining. No significant correlation between size and accuracy ($p = 0.24$) confirms that scaling beyond 28M parameters yields no systematic improvement. \textbf{(B) AUROC vs Parameter Count}: Pretrained models achieve exceptional 0.9633--0.9997, saturating around 28M. Scratch models 0.8537--0.9832 show high variance. Vertical separation larger than OCT, reflecting stronger ImageNet transfer to RGB CFP. No significant correlation ($p = 0.29$). \textbf{(C) F1-Score vs Parameter Count}: Mirrors accuracy; pretrained 0.93--0.99, scratch 0.79--0.95. No significant correlation ($p = 0.24$). Consistent message: compact pretrained models (27--29M) achieve near-perfect performance; scaling to 300M yields no benefit for this CFP 3-class task.}
\label{fig:performance_size_dme}
\end{figure*}

\begin{figure*}[!ht]
\centering
\includegraphics[width=0.85\textwidth]{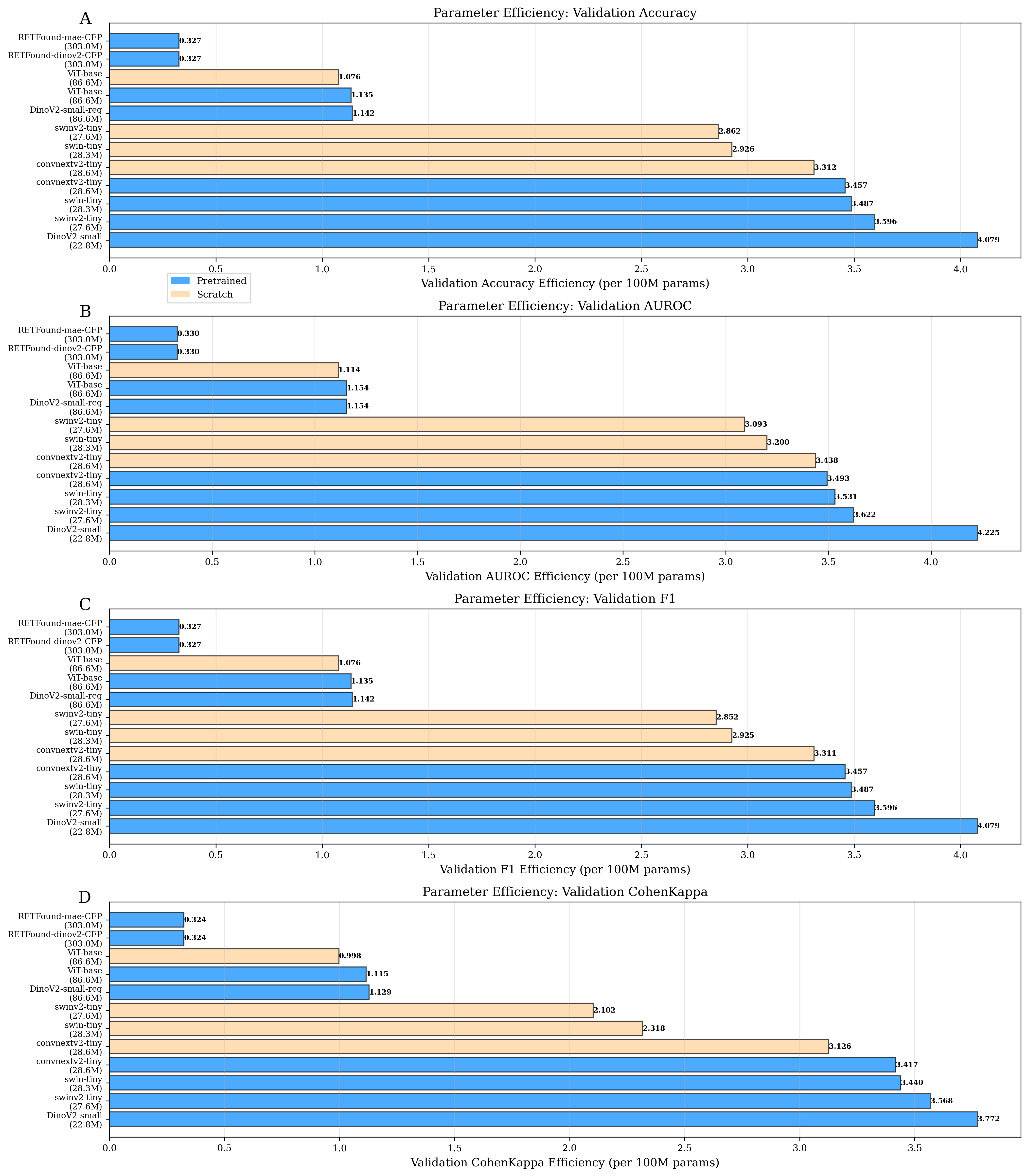}
\caption{DME: Multi-panel horizontal bar chart showing parameter efficiency across all 12 models. Y-axis: model names with parameter counts; X-axis: efficiency [0.0--5.0]. \textbf{(A) Accuracy Efficiency}: DinoV2-small leads with 4.079 (93.01\%, 22.8M), excellent for resource-constrained deployment. However, SwinV2-tiny (3.596, 99.24\%, 27.6M) provides significant 6.23 percentage point gain for modest parameter increase. RETFound-MAE-CFP shows poor efficiency (0.327, 99.06\%, 303M), 10.9$\times$ lower than DinoV2-small for similar performance. Scratch models <3.5 due to lower accuracy. \textbf{(B) AUROC Efficiency}: DinoV2-small leads (4.225), pretrained models <4.2, scratch <3.4. \textbf{(C) F1 Efficiency}: Similar pattern; DinoV2-small (4.079), compact pretrained <3.3, RETFound 0.33. \textbf{(D) Kappa Efficiency}: Pretrained <3.78, scratch models showing lower efficiency (<3.13) reflecting poor chance-adjusted agreement. Consistent message: compact pretrained models deliver superior efficiency; oversized specialized models show poor parameter utilization on this task.}
\label{fig:efficiency_dme}
\end{figure*}

\begin{figure*}[!ht]
\centering
\includegraphics[width=1.0\textwidth]{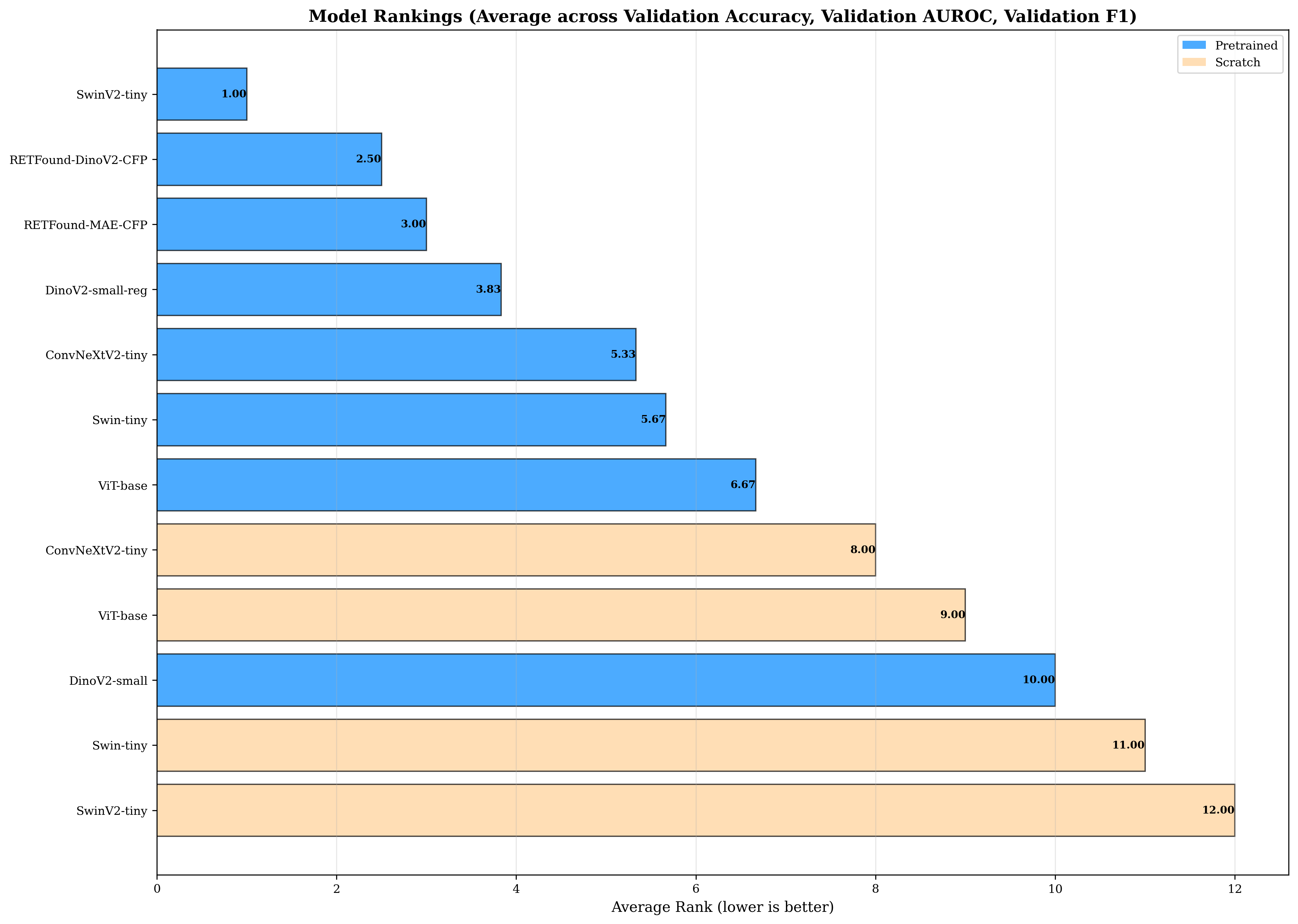}
\caption{DME: Aggregate model rankings across accuracy, AUROC, and F1-score for diabetic macular edema classification. SwinV2-tiny achieves perfect average rank (1.00), ranking first on all three metrics, the only model in the study to achieve this distinction. The ranking stratification is near absolute: pretrained models occupy ranks 1--7, scratch-trained models occupy ranks 8--12, with one pretrained overlap in rank 10. RETFound domain-specific models rank 2nd and 3rd (DinoV2-CFP: 2.50, MAE-CFP: 3.00), demonstrating competitive but not superior performance compared to the compact general-purpose SwinV2-tiny. This ranking pattern indicates that for 3-class grading with clear visual markers, ImageNet-pretrained hierarchical architectures in the 27--29M parameter range provide optimal overall performance without requiring specialized retinal foundation models.}
\label{fig:rankings_dme}
\end{figure*}

All pretrained models (ranks 1--8) outperformed all scratch-trained models (ranks 9--12), demonstrating consistent pretraining advantage (Figure~\ref{fig:rankings_dme}). The DinoV2-small achieved the highest parameter efficiency (4.08 per 100M), delivering 93.01\% accuracy with only 22.8M parameters (Figure~\ref{fig:efficiency_dme}). Pareto frontier analysis identified two optimal models: dinov2-small (22.8M, 93.01\% accuracy) and SwinV2-tiny (27.6M, 99.24\% accuracy), as illustrated in Figure~\ref{fig:pareto_dme}. Figure~\ref{fig:performance_size_dme} shows the performance-size trade-offs across all models. Pearson correlation analysis revealed no significant linear relationship between model size and performance (accuracy: $p = 0.24$; AUROC: $p = 0.29$; F1-score: $p = 0.24$), confirming that increased parameter count does not predict higher accuracy for this DME classification task.

\subsubsection{Architecture-Specific Observations}
Hierarchical architectures (Swin Transformers, ConvNeXt) achieved the highest absolute performance when pretrained. Domain-specific RETFound models reached 99.06\% accuracy but were outperformed by the smaller Swin Transformers, suggesting ImageNet pretraining transfers effectively. The pretraining benefit varied dramatically by architecture: SwinV2-tiny improved by 20.25 percentage points (99.24\% pretrained vs.\ 78.99\% scratch), while ConvNeXtV2 showed only 4.16 percentage points improvement. See Appendix~\ref{app:dme_figures} for additional visualizations.

\subsection{Diabetic Retinopathy (DR) Classification Results from CFP data}
\label{sec:results:cfp_dr}

\subsubsection{Overall Performance}
\begin{figure*}[!ht]
\centering
\includegraphics[width=0.85\textwidth]{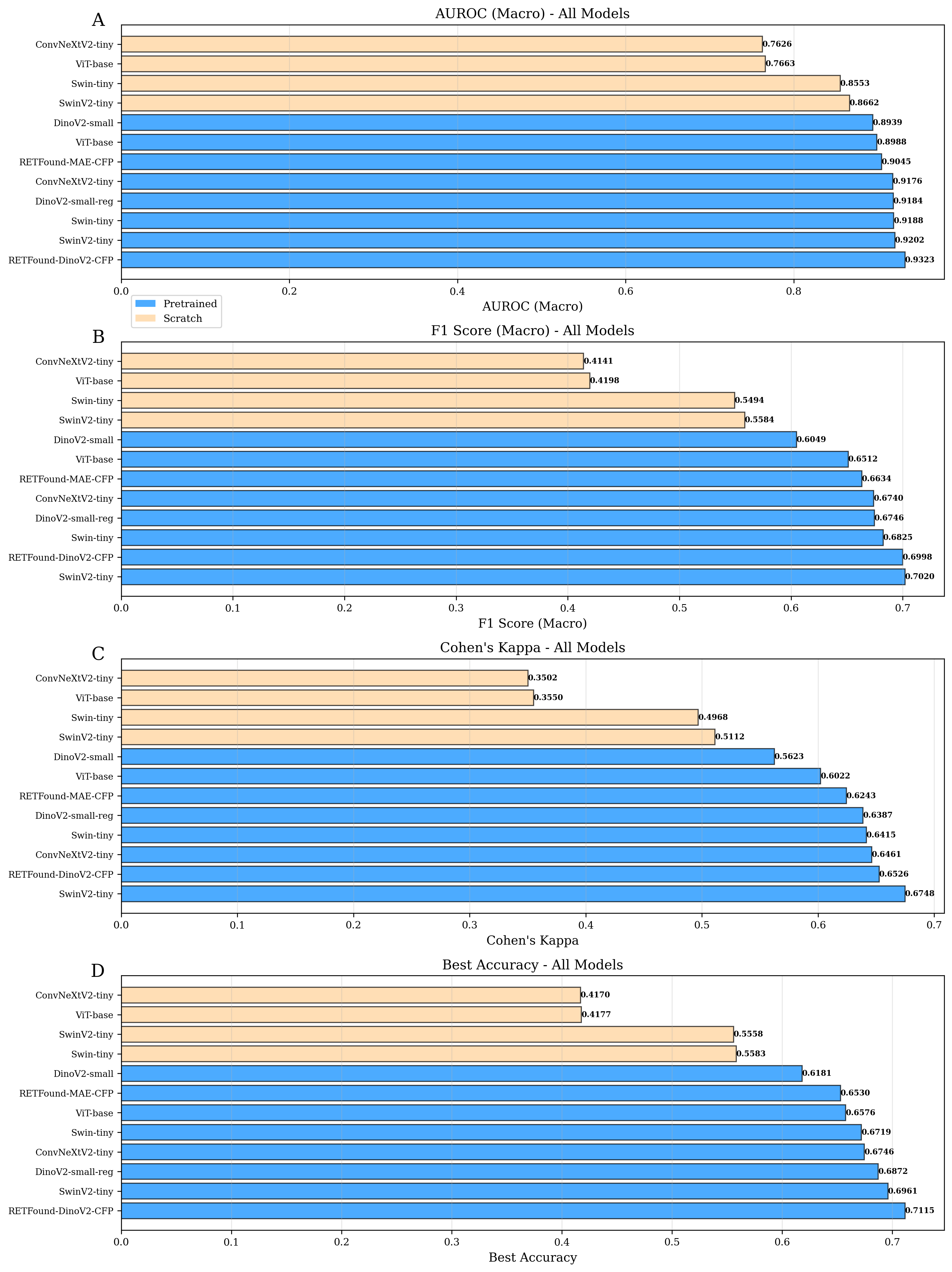}
\caption{DR: Multi-panel horizontal bar chart for diabetic retinopathy severity grading, the only task where domain-specific models achieve top performance. Models on Y-axis (sorted by performance), values [0.0--1.0] on X-axis. \textbf{(A) AUROC (macro)}: RETFound-DinoV2-CFP leads at 0.9323, with pretrained models 0.8939--0.9323, scratch 0.7626--0.8662. Substantially lower than other tasks, reflecting grading difficulty. \textbf{(B) F1-Score (macro)}: SwinV2-tiny achieves highest F1 (0.7020), with RETFound-DinoV2 at 0.6998. Pretrained 0.6049--0.7020, scratch 0.4141--0.5584. The compressed range (0.40--0.70) demonstrates task difficulty. \textbf{(C) Cohen's Kappa}: Pretrained 0.5623--0.6748, scratch 0.3502--0.5112. Lower Kappa values reflect challenge of chance-adjusted agreement under severe class imbalance (Class 3: 8.22\%). \textbf{(D) Accuracy}: RETFound-DinoV2-CFP achieves 0.7115, outperforming ImageNet-pretrained SwinV2 (0.6961) by 1.54 percentage points, the only task where retina-specific pretraining provides measurable advantage. All absolute values (0.6181--0.7115 pretrained, 0.4170--0.5583 scratch) substantially lower than OCT/DME/GL, confirming this as the most challenging task.}
\label{fig:validation_metrics_dr}
\end{figure*}

Diabetic retinopathy severity grading proved substantially more challenging than other tasks. We evaluated 12 model configurations (22.8M--303M parameters), with the best model, RETFound-DinoV2-CFP (domain-specific, 303M parameters), achieving only 71.15\% validation accuracy, 93.23\% AUROC, and 69.98\% F1-score. This represents a 26 percentage point decrease compared to OCT performance, reflecting the difficulty of fine-grained 5-class ordinal grading on the highly imbalanced EAM dataset. Table~\ref{tab:top_models_dr} summarizes the top-performing models. Figure~\ref{fig:validation_metrics_dr} provides a comprehensive comparison of validation metrics across all 12 models.

\begin{table*}[t]
\centering
\caption{Top-5 performing models on DR classification. All top models use pretrained initialization. Notably, the domain-specific RETFound model achieved the highest accuracy on this challenging 5-class severity grading task. Metrics as defined in Table~\ref{tab:top_models}.}
\label{tab:top_models_dr}
\small
\begin{tabular}{@{}lcccccc@{}}
\toprule
\textbf{Model} & \textbf{Pretrained} & \textbf{Params} & \textbf{Acc (\%)} & \textbf{AUROC} & \textbf{F1} & \textbf{Avg. Rank} \\
\midrule
RETFound-DinoV2-CFP & Yes & 303M & 71.15 & 0.9323 & 0.6998 & 1.33 \\
SwinV2-tiny & Yes & 27.6M & 69.61 & 0.9202 & 0.7020 & 1.67 \\
Swin-tiny & Yes & 28.3M & 67.19 & 0.9188 & 0.6825 & 3.67 \\
DinoV2-small-reg & Yes & 86.6M & 68.72 & 0.9184 & 0.6746 & 3.67 \\
ConvNeXtV2-tiny & Yes & 28.6M & 67.46 & 0.9176 & 0.6740 & 4.67 \\
\bottomrule
\end{tabular}
\end{table*}

\subsubsection{Impact of Pretraining}
\begin{figure*}[!ht]
\centering
\includegraphics[width=1.0\textwidth]{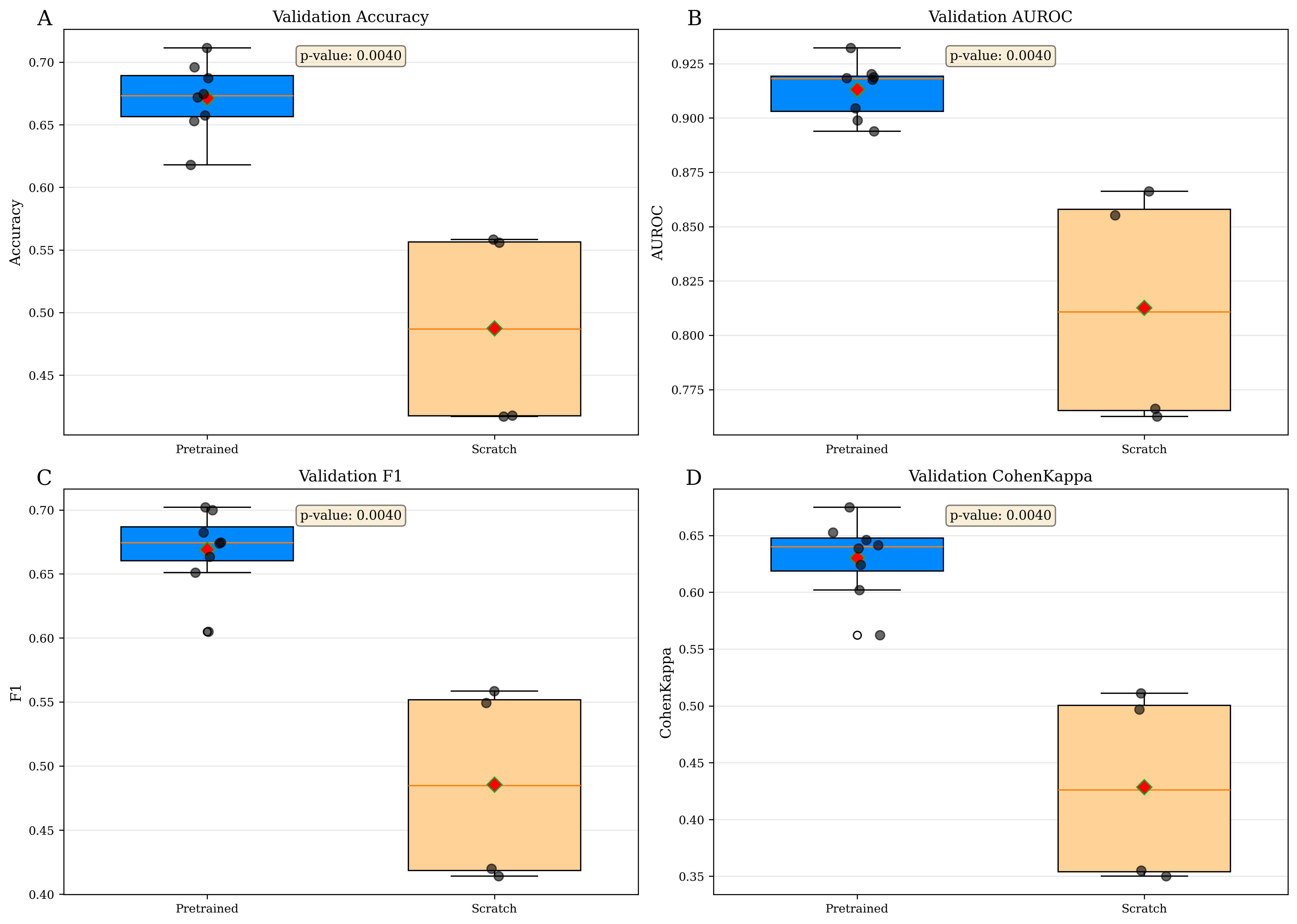}
\caption{DR: Multi-panel box plot comparison for diabetic retinopathy severity grading, the most challenging task in the study. \textbf{(A) Best Validation Accuracy}: Pretrained models (n=8) achieve mean 67.13\% $\pm$ 2.89\% vs scratch (n=4) 48.72\% $\pm$ 8.07\%, representing an 18.41 percentage point improvement (p < 0.05), the largest pretraining benefit across all tasks, exceeding OCT (5.18\%) by 3.6$\times$. This substantial advantage indicates pretrained features become especially valuable for challenging tasks requiring fine-grained discrimination under severe class imbalance (Class 3: 8.22\%). The 2.8$\times$ variance reduction demonstrates critical importance for training stability. \textbf{(B) AUROC (macro)}: Pretrained 91.31\% $\pm$ 1.28\% vs scratch 81.26\% $\pm$ 5.58\%, a 10.05 percentage point improvement (p < 0.05), showing that pretraining substantially improves class discrimination. \textbf{(C) F1-Score (macro)}: Pretrained 66.90\% $\pm$ 3.10\% vs scratch 48.54\% $\pm$ 7.92\%, an 18.36 percentage point improvement. \textbf{(D) Cohen's Kappa}: Pretrained 63.03\% $\pm$ 3.46\% vs scratch 42.83\% $\pm$ 8.76\%, showing a 20.20 percentage point improvement, critical for imbalanced tasks where chance agreement is substantial.}
\label{fig:pretrained_vs_scratch_dr}
\end{figure*}

Pretraining remained critically important for DR despite the overall lower performance (Table~\ref{tab:pretrained_comparison_dr}, Figure~\ref{fig:pretrained_vs_scratch_dr}). Pretrained models achieved mean best validation accuracy of $0.6713 \pm 0.0289$ (n=8) versus $0.4872 \pm 0.0807$ (n=4) for scratch models, representing an absolute improvement of 18.41 percentage points (p < 0.05). This pretraining advantage (18.41\%) substantially exceeds that observed for OCT (5.18\%) and DME (10.70\%), indicating pretraining is especially valuable for fine-grained ordinal classification tasks with class imbalance.

\begin{table*}[t]
\centering
\caption{Comparison of pretrained vs.\ scratch-trained models for DR. The large pretraining benefit reflects the difficulty of ordinal severity grading on imbalanced data. Metrics as defined in Table~\ref{tab:pretrained_comparison}.}
\label{tab:pretrained_comparison_dr}
\small
\begin{tabular}{@{}lcccc@{}}
\toprule
\textbf{Metric} & \textbf{Pretrained (n=8)} & \textbf{Scratch (n=4)} & \textbf{Difference} & \textbf{p-value} \\
\midrule
AUROC (Macro) & 0.9131 $\pm$ 0.0128 & 0.8126 $\pm$ 0.0558 & +10.05\% & 0.0040 \\
F1-Score (Macro) & 0.6690 $\pm$ 0.0310 & 0.4854 $\pm$ 0.0792 & +18.36\% & 0.0040 \\
Cohen's Kappa & 0.6303 $\pm$ 0.0346 & 0.4283 $\pm$ 0.0876 & +20.20\% & 0.0040 \\
Best Validation Acc & 0.6713 $\pm$ 0.0289 & 0.4872 $\pm$ 0.0807 & +18.41\% & 0.0040 \\
\bottomrule
\end{tabular}
\end{table*}

\subsubsection{Model Rankings and Efficiency}
\begin{figure*}[!ht]
\centering
\includegraphics[width=1.0\textwidth]{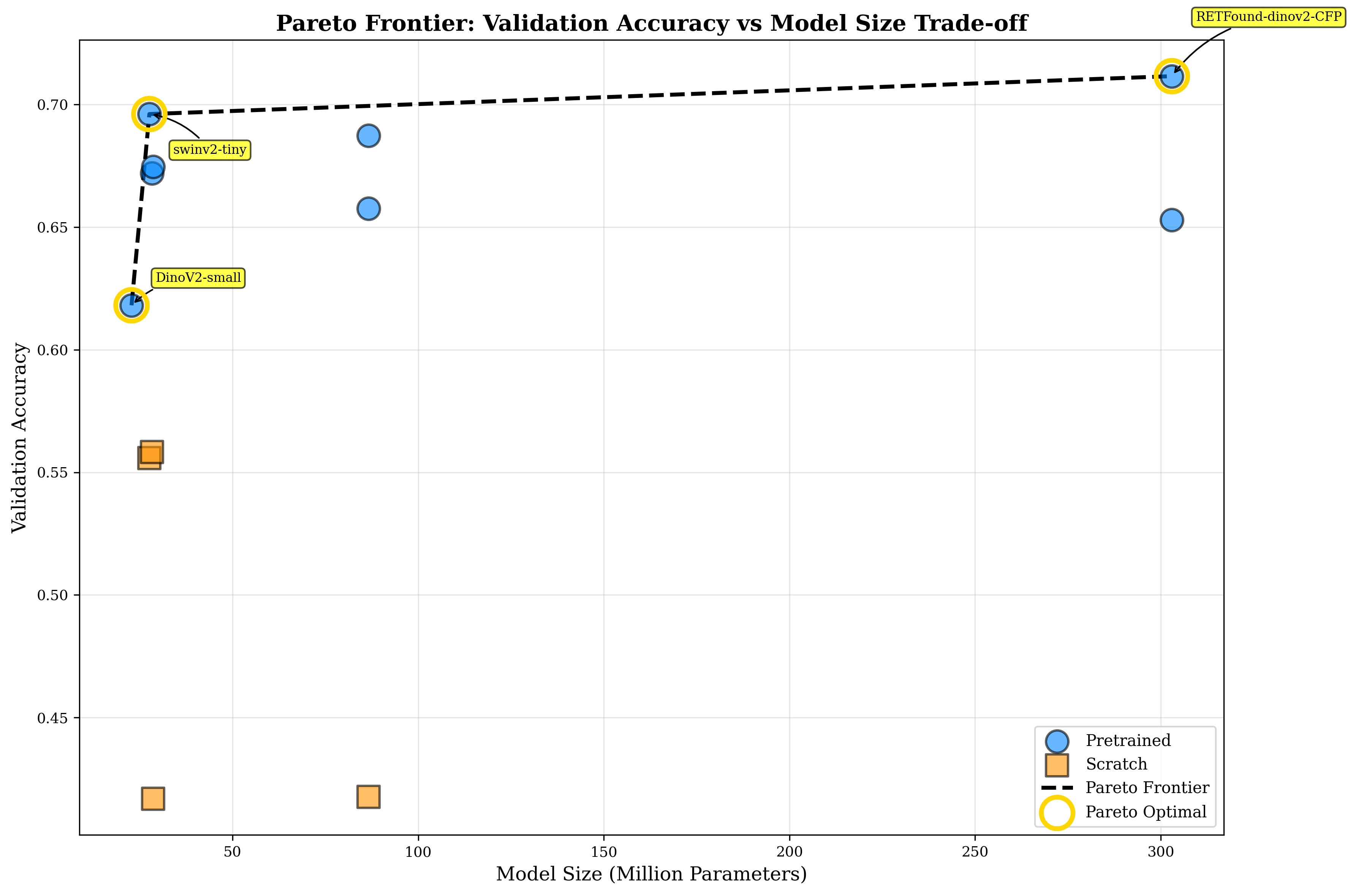}
\caption{DR: Pareto frontier analysis for diabetic retinopathy severity grading. Unlike other tasks where compact models (23--29M) dominated, the DR frontier spans a wider range from DinoV2-small (22.8M, 61.81\%) to RETFound-DinoV2-CFP (303M, 71.15\%), with SwinV2-tiny (27.6M, 69.61\%) occupying an intermediate position. This is the only task where the 303M domain-specific model achieves Pareto optimality, justifying its 11$\times$ increased size relative to SwinV2 through a 1.54 percentage point accuracy gain. The frontier's broader parameter range suggests that task difficulty modulates the optimal model size: tractable tasks (DME, OCT) saturate at 28M parameters, while challenging grading with severe class imbalance benefits from larger specialized models. This defines the boundary case where computational investment in domain-specific 300M+ parameter foundation models provides measurable, albeit modest, performance advantages.}
\label{fig:pareto_dr}
\end{figure*}

\begin{figure*}[!ht]
\centering
\includegraphics[width=1.0\textwidth]{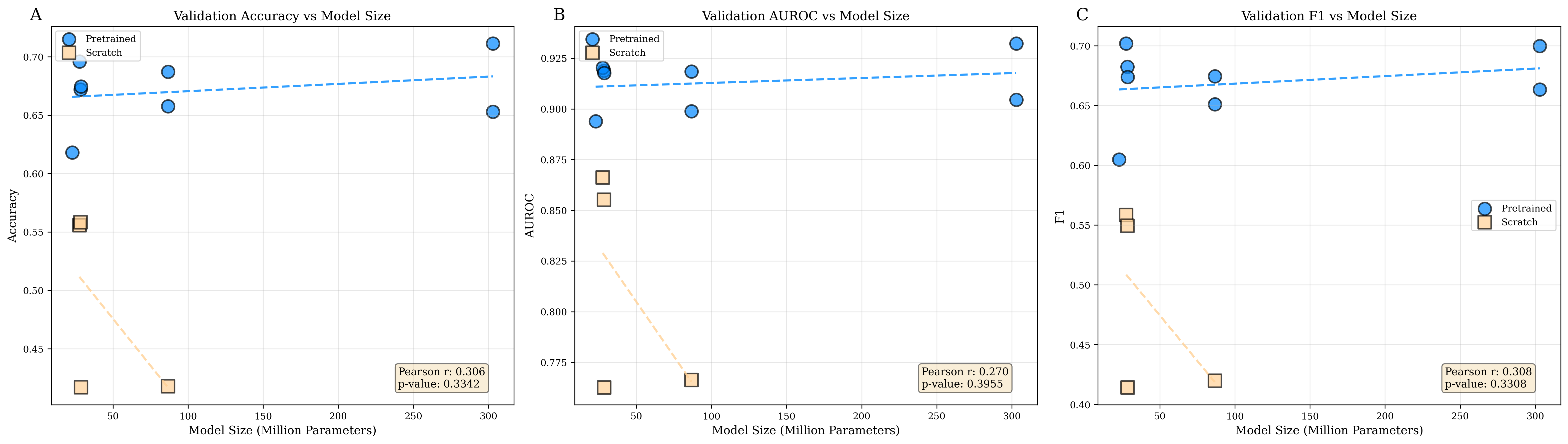}
\caption{DR: Multi-panel scatter plots for diabetic retinopathy showing the largest pretraining advantage (18.41 percentage points) across all tasks. Pretrained (blue), scratch (green). Pearson correlation statistics (r, p-value) quantify the linear relationship between model size and performance. \textbf{(A) Accuracy vs Parameter Count}: Substantial vertical separation; pretrained cluster 0.61--0.71, scratch 0.42--0.58, more than 3$\times$ larger gap than OCT, demonstrating pretraining benefits scale with task difficulty. Unlike OCT/DME where performance plateaus at 28M, DR shows continued modest gains: SwinV2-tiny (27.6M, 0.6961) < RETFound (303M, 0.7115), suggesting specialized large models provide value for challenging classification tasks. However, no significant correlation between size and accuracy ($p = 0.33$) indicates that model scale alone does not predict performance even for this challenging task. Compressed accuracy range (0.42--0.71) reflects extreme difficulty of the 5-class severity grading under imbalance. \textbf{(B) AUROC vs Parameter Count}: Pretrained 0.8939--0.9323, scratch 0.7626--0.8662. No significant correlation ($p = 0.40$). \textbf{(C) F1-Score vs Parameter Count}: Pretrained 0.6049--0.7020, scratch 0.4141--0.5584. No significant correlation ($p = 0.33$). Consistent message: this is the only task where the largest model (303M RETFound) achieves top performance, yet the lack of linear correlation shows that architecture and pretraining type matter more than raw parameter count.}
\label{fig:performance_size_dr}
\end{figure*}

\begin{figure*}[!ht]
\centering
\includegraphics[width=0.85\textwidth]{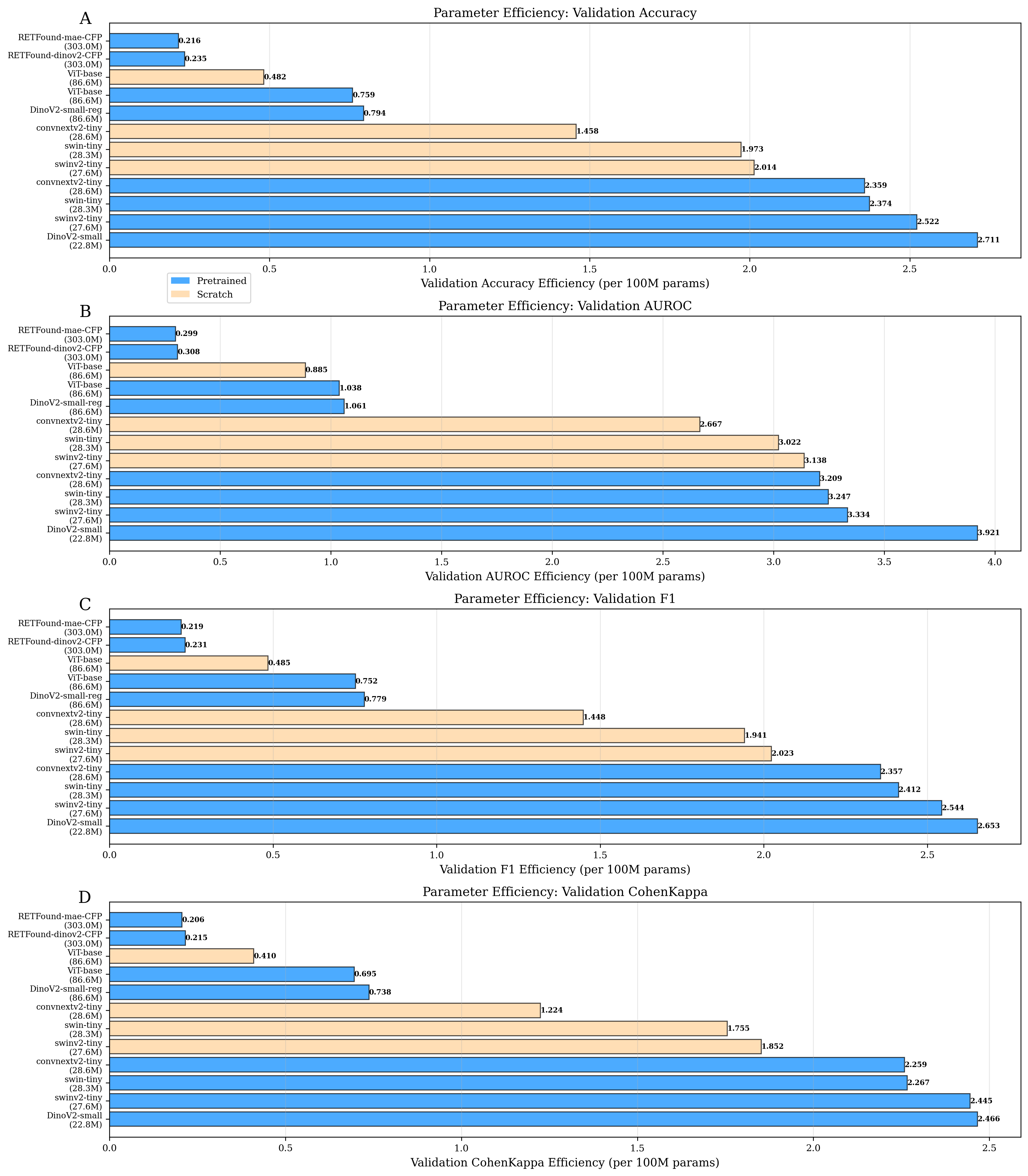}
\caption{DR: Multi-panel horizontal bar chart showing parameter efficiency for diabetic retinopathy severity grading. Y-axis: models with sizes; X-axis: efficiency [0.0--5.0]. \textbf{(A) Accuracy Efficiency}: DinoV2-small achieves 2.71 (61.81\%, 22.8M), substantially lower than its 4.20 on OCT. A trade-off is observed: compact models show superior efficiency but lower absolute accuracy (DinoV2-small: 61.81\%) versus specialized large models (RETFound: 71.15\%, 0.23 efficiency). \textbf{(B) AUROC Efficiency}: DinoV2-small (3.921), pretrained <3.921, RETFound (<0.308). \textbf{(C) F1 Efficiency}: DinoV2-small (2.653), pretrained <2.653, RETFound (<0.231). \textbf{(D) Kappa Efficiency}: DinoV2-small (2.466), pretrained <2.466, RETFound (<0.215). Unlike DME/OCT where compact models achieve near-optimal absolute performance, DR illustrates that optimal model selection must account for task difficulty.}
\label{fig:efficiency_dr}
\end{figure*}

\begin{figure*}[!ht]
\centering
\includegraphics[width=1.0\textwidth]{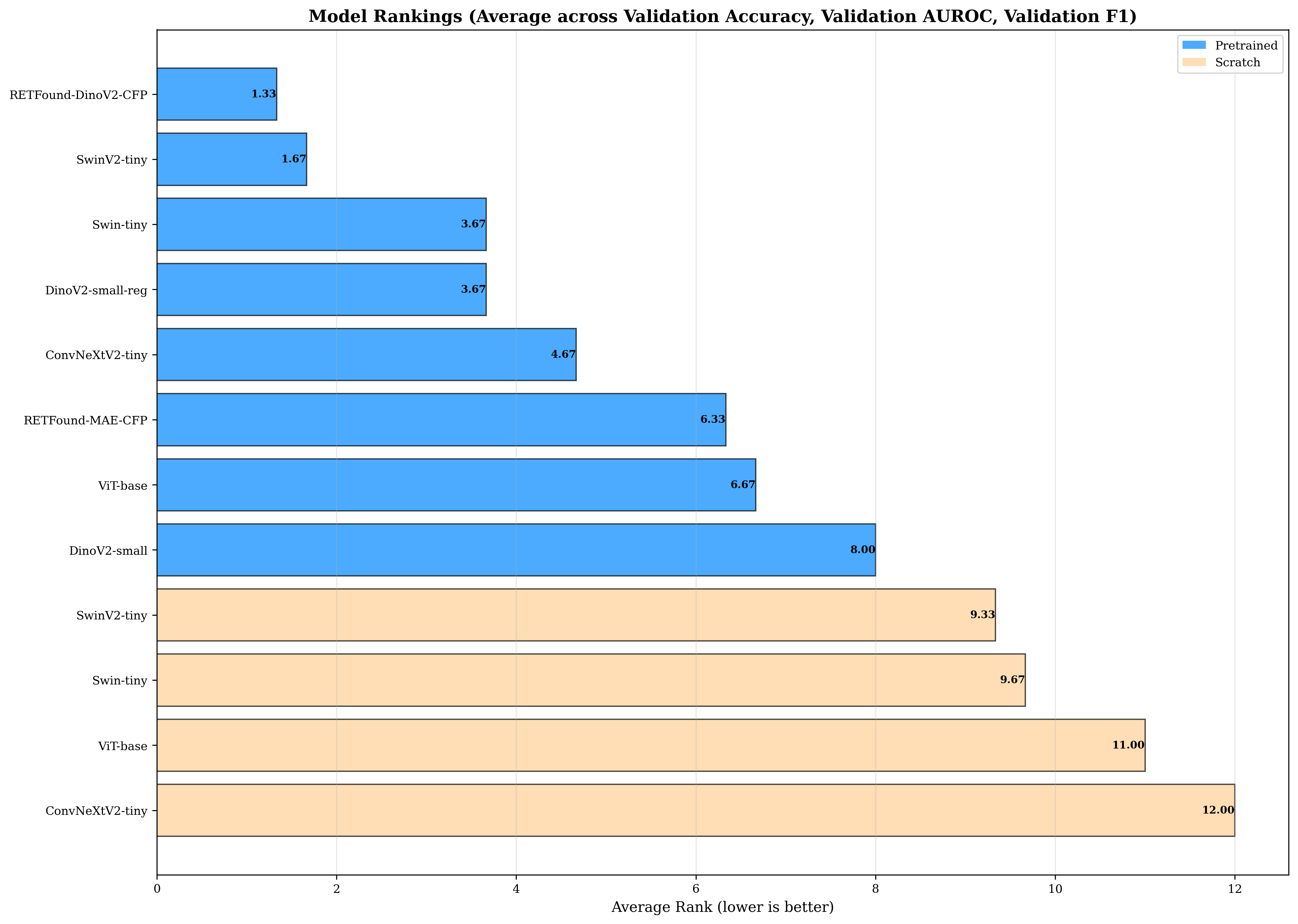}
\caption{DR: Aggregate model rankings for diabetic retinopathy severity grading. RETFound-DinoV2-CFP achieves the best average rank (1.33), marking the only task where domain-specific retinal pretraining outperforms general-purpose ImageNet initialization. SwinV2-tiny ranks second (1.67), demonstrating competitive performance at 11$\times$ smaller size. The ranking stratification remains absolute: all pretrained models (ranks 1--8) outperform all scratch-trained models (ranks 9--12). Notably, the compact DinoV2-small ranks 8th among pretrained models (vs 5th--6th on other tasks), indicating that small architectures struggle more. This ranking pattern reveals the boundary case where specialized foundation models justify their computational cost: only for fine-grained severity discrimination under extreme class imbalance do 300M+ parameter retina-specific models provide demonstrable advantages over compact general-purpose alternatives.}
\label{fig:rankings_dr}
\end{figure*}

Unlike OCT and DME, domain-specific RETFound-DinoV2-CFP achieved the highest accuracy (71.15\%) on DR, outperforming ImageNet-pretrained models. This marks the only task where retina-specific pretraining provided clear advantages over general-purpose ImageNet initialization. All pretrained models (ranks 1--8) again outperformed scratch-trained models (ranks 9--12), as shown in Figure~\ref{fig:rankings_dr}. Parameter efficiency analysis identified DinoV2-small (61.81\%, 22.8M, efficiency 2.71) as optimal for resource-constrained deployments (Figure~\ref{fig:efficiency_dr}). Pareto frontier included three models spanning 22.8M to 303M parameters (Figure~\ref{fig:pareto_dr}), with performance-size trade-offs illustrated in Figure~\ref{fig:performance_size_dr}. Pearson correlation analysis revealed no significant linear relationship between model size and performance (accuracy: $p = 0.33$; AUROC: $p = 0.40$; F1-score: $p = 0.33$), indicating that even for this challenging task, model scale alone does not determine performance.

\subsubsection{Architecture-Specific Observations}
Hierarchical Swin Transformers maintained strong performance (Swin-tiny: 67.19\%, SwinV2-tiny: 69.61\%), with SwinV2-tiny ranking second overall despite 11$\times$ fewer parameters than RETFound. The pretraining benefit was substantial but varied by architecture: SwinV2 improved by 14.03 percentage points (69.61\% vs 55.58\% scratch), while ViT-base showed an even larger 23.99 percentage points improvement (65.76\% vs 41.77\% scratch). The consistently lower absolute performance across all models (compared to OCT/DME) reflects the inherent difficulty of distinguishing subtle ordinal differences in DR severity levels with extreme class imbalance (Class 3: 8.22\%). See Appendix~\ref{app:dr_figures} for additional visualizations.

\subsection{Glaucoma (GL) Classification Results from CFP data}
\label{sec:results:cfp_gl}

\subsubsection{Overall Performance}
\begin{figure*}[!ht]
\centering
\includegraphics[width=0.85\textwidth]{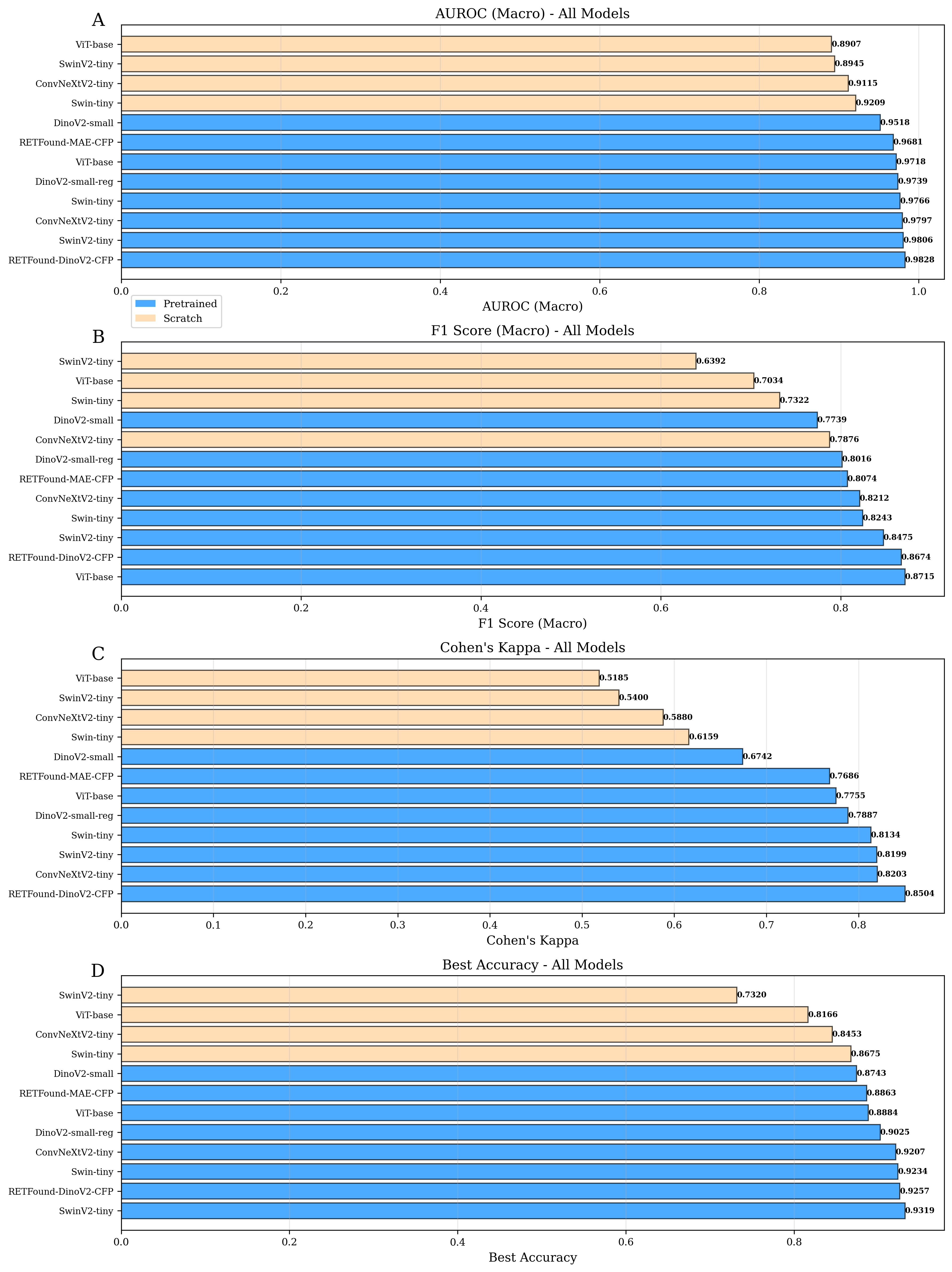}
\caption{GL: Multi-panel horizontal bar chart for glaucoma detection on combined AIROGS + PAPILA datasets. Models on Y-axis (sorted by performance), values [0.0--1.0] on X-axis. \textbf{(A) AUROC (macro)}: RETFound-DinoV2-CFP leads at 0.9828, with pretrained models clustering 0.9518--0.9828, scratch 0.8907--0.9209. All pretrained models exceed 0.95, demonstrating excellent threshold-independent discrimination for this clinical screening application. Performance falls between DME (very high, >0.99) and DR (challenging, 0.89--0.93). \textbf{(B) F1-Score (macro)}: ViT-base achieves highest F1 (0.8715), with pretrained 0.7739--0.8715, scratch 0.6392--0.7876. Wider F1 range reflects class imbalance in 3-class labels. \textbf{(C) Cohen's Kappa}: Pretrained 0.6742--0.8504, scratch 0.5185--0.6159. Moderate Kappa values reflect 3-class task complexity. \textbf{(D) Accuracy}: SwinV2-tiny leads at 0.9319, marginally outperforming 303M RETFound-DinoV2-CFP (0.9257) by 0.62 percentage points despite 11$\times$ fewer parameters. This demonstrates domain-specific retinal pretraining provides no measurable benefit over ImageNet initialization for this task.}
\label{fig:validation_metrics_gl}
\end{figure*}

We evaluated 12 model configurations on glaucoma detection across the combined AIROGS and PAPILA datasets (22.8M--303M parameters). The highest accuracy was achieved by Swin-tiny (pretrained, 27.6M parameters) at 93.19\% validation accuracy, 98.06\% AUROC, and 84.75\% F1-score. Performance fell between DME (very high) and DR (challenging), reflecting moderate task difficulty in this 3-class glaucoma detection task. Table~\ref{tab:top_models_gl} summarizes the top-performing models. Figure~\ref{fig:validation_metrics_gl} provides a comprehensive comparison of validation metrics across all 12 models.

\begin{table*}[t]
\centering
\caption{Top-5 performing models on GL classification. All top models use pretrained initialization. Models were evaluated on the combined AIROGS + PAPILA dataset with 3-class labels. Metrics as defined in Table~\ref{tab:top_models}.}
\label{tab:top_models_gl}
\small
\begin{tabular}{@{}lcccccc@{}}
\toprule
\textbf{Model} & \textbf{Pretrained} & \textbf{Params} & \textbf{Acc (\%)} & \textbf{AUROC} & \textbf{F1} & \textbf{Avg. Rank} \\
\midrule
RETFound-DinoV2-CFP & Yes & 303M & 92.57 & 0.9828 & 0.8674 & 1.67 \\
SwinV2-tiny & Yes & 27.6M & 93.19 & 0.9806 & 0.8475 & 2.00 \\
Swin-tiny & Yes & 28.3M & 92.34 & 0.9766 & 0.8243 & 3.67 \\
ConvNeXtV2-tiny & Yes & 28.6M & 92.07 & 0.9797 & 0.8212 & 4.00 \\
ViT-base & Yes & 86.6M & 88.84 & 0.9718 & 0.8715 & 4.33 \\
\bottomrule
\end{tabular}
\end{table*}

\subsubsection{Impact of Pretraining}
\begin{figure*}[!ht]
\centering
\includegraphics[width=1.0\textwidth]{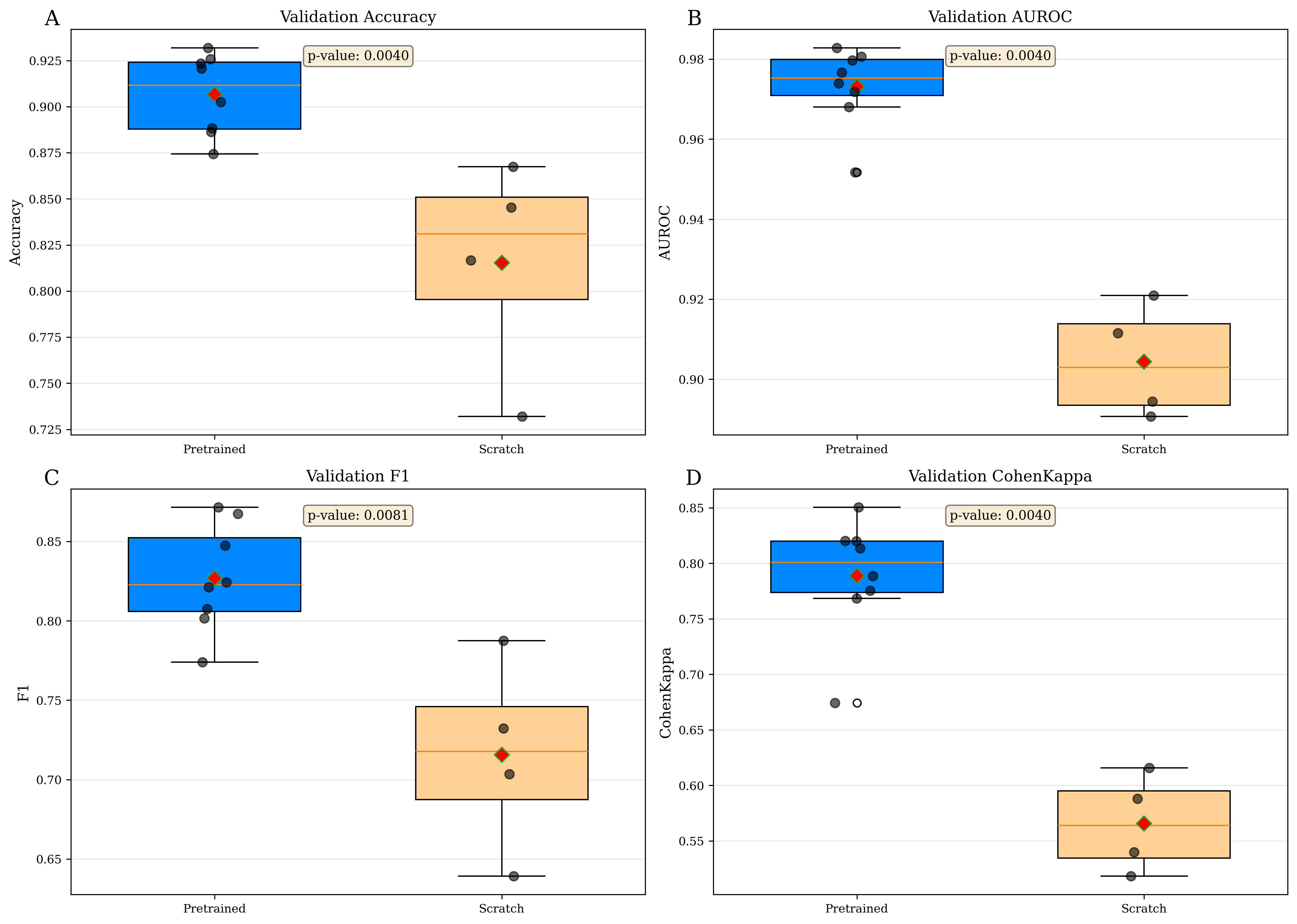}
\caption{GL: Multi-panel box plot comparison for glaucoma detection on combined AIROGS + PAPILA datasets (114,381 images, 3-class labels). \textbf{(A) Best Validation Accuracy}: Pretrained models (n=8) achieve mean 90.66\% $\pm$ 2.17\% vs scratch (n=4) 81.54\% $\pm$ 5.93\%, representing a 9.13 percentage point improvement (p < 0.05). This pretraining advantage falls between OCT (5.18\%) and the more challenging DME (10.70\%) and DR (18.41\%), consistent with moderate task difficulty. The 2.7$\times$ variance reduction demonstrates improved training stability. Despite the large AIROGS dataset (113,893 images), pretrained models still significantly outperform scratch training, indicating ImageNet features transfer effectively even with abundant task-specific data. \textbf{(B) AUROC (macro)}: Pretrained 97.32\% $\pm$ 0.99\% vs scratch 90.44\% $\pm$ 1.43\%, a 6.88 percentage point improvement (p < 0.05). \textbf{(C) F1-Score (macro)}: Pretrained 82.68\% $\pm$ 3.36\% vs scratch 71.56\% $\pm$ 6.17\%, an 11.12 percentage point improvement (p < 0.05). \textbf{(D) Cohen's Kappa}: Pretrained 78.89\% $\pm$ 5.36\% vs scratch 56.56\% $\pm$ 4.44\%, showing a 22.33 percentage point improvement—the largest Kappa benefit across all tasks, indicating pretraining is especially valuable for chance-adjusted agreement in this 3-class glaucoma detection task.}
\label{fig:pretrained_vs_scratch_gl}
\end{figure*}

Pretrained models substantially outperformed scratch-trained models on glaucoma detection (Table~\ref{tab:pretrained_comparison_gl}, Figure~\ref{fig:pretrained_vs_scratch_gl}). Pretrained models achieved mean best validation accuracy of $0.9066 \pm 0.0217$ (n=8) versus $0.8154 \pm 0.0593$ (n=4) for scratch models, representing an absolute improvement of 9.13 percentage points (p < 0.05). This pretraining advantage (9.13\%) falls between OCT (5.18\%) and the more challenging DME (10.70\%) and DR (18.41\%) tasks, suggesting that pretraining provides greater value for CFP-based tasks compared to OCT imaging modality, with the benefit scaling with task difficulty. Similar patterns were observed for AUROC (pretrained: $0.9732 \pm 0.0099$ vs.\ scratch: $0.9044 \pm 0.0143$) and F1-score (pretrained: $0.8268 \pm 0.0336$ vs.\ scratch: $0.7156 \pm 0.0617$).

\begin{table*}[t]
\centering
\caption{Comparison of pretrained vs.\ scratch-trained models for GL. The large pretraining benefit reflects the value of ImageNet features for CFP-based glaucoma detection. Metrics as defined in Table~\ref{tab:pretrained_comparison}.}
\label{tab:pretrained_comparison_gl}
\small
\begin{tabular}{@{}lcccc@{}}
\toprule
\textbf{Metric} & \textbf{Pretrained (n=8)} & \textbf{Scratch (n=4)} & \textbf{Difference} & \textbf{p-value} \\
\midrule
AUROC (Macro) & 0.9732 $\pm$ 0.0099 & 0.9044 $\pm$ 0.0143 & +6.88\% & 0.0040 \\
F1-Score (Macro) & 0.8268 $\pm$ 0.0336 & 0.7156 $\pm$ 0.0617 & +11.12\% & 0.0081 \\
Cohen's Kappa & 0.7889 $\pm$ 0.0536 & 0.5656 $\pm$ 0.0444 & +22.33\% & 0.0040 \\
Best Validation Acc & 0.9066 $\pm$ 0.0217 & 0.8154 $\pm$ 0.0593 & +9.13\% & 0.0040 \\
\bottomrule
\end{tabular}
\end{table*}

\subsubsection{Model Rankings and Efficiency}
\begin{figure*}[!ht]
\centering
\includegraphics[width=1.0\textwidth]{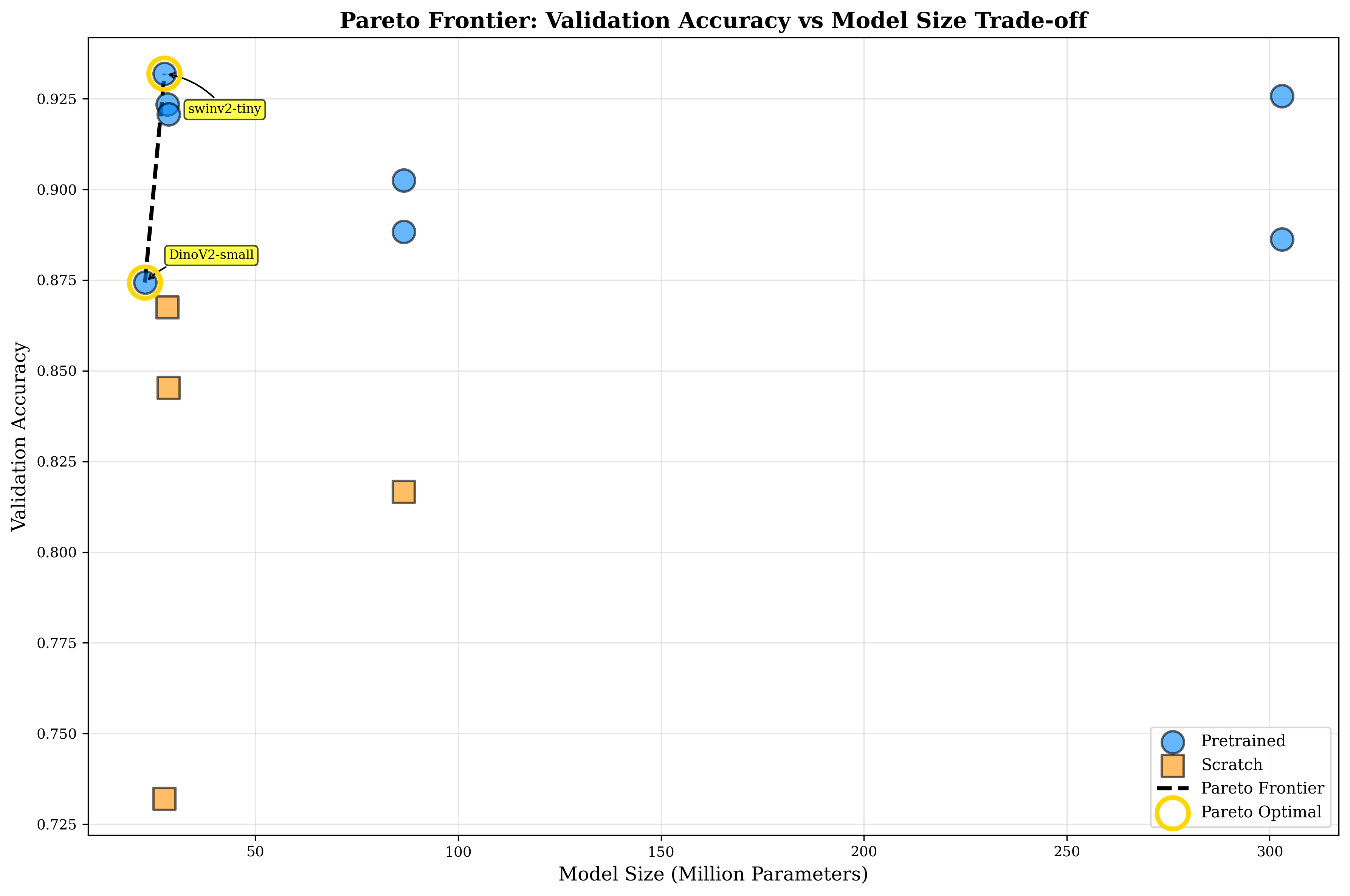}
\caption{GL: Pareto frontier analysis for glaucoma detection. Two models achieve Pareto optimality: DinoV2-small (22.8M, 87.43\% accuracy) and SwinV2-tiny (27.6M, 93.19\% accuracy). Similar to DME but unlike DR, the frontier is dominated exclusively by compact models in the 23--29M parameter range; the 303M RETFound model (92.57\%) falls below the frontier, dominated by the smaller SwinV2-tiny which achieves higher accuracy at 11$\times$ lower parameter cost. The steep frontier slope (5.76 percentage point gain for 4.8M additional parameters between the two frontier models) indicates that modest scaling from 23M to 28M provides substantial performance benefits, but further scaling to 303M yields negative returns. This pattern reinforces the finding that domain-specific foundation models are unnecessary for moderate-difficulty CFP tasks with clear visual features and abundant training data.}
\label{fig:pareto_gl}
\end{figure*}

\begin{figure*}[!ht]
\centering
\includegraphics[width=1.0\textwidth]{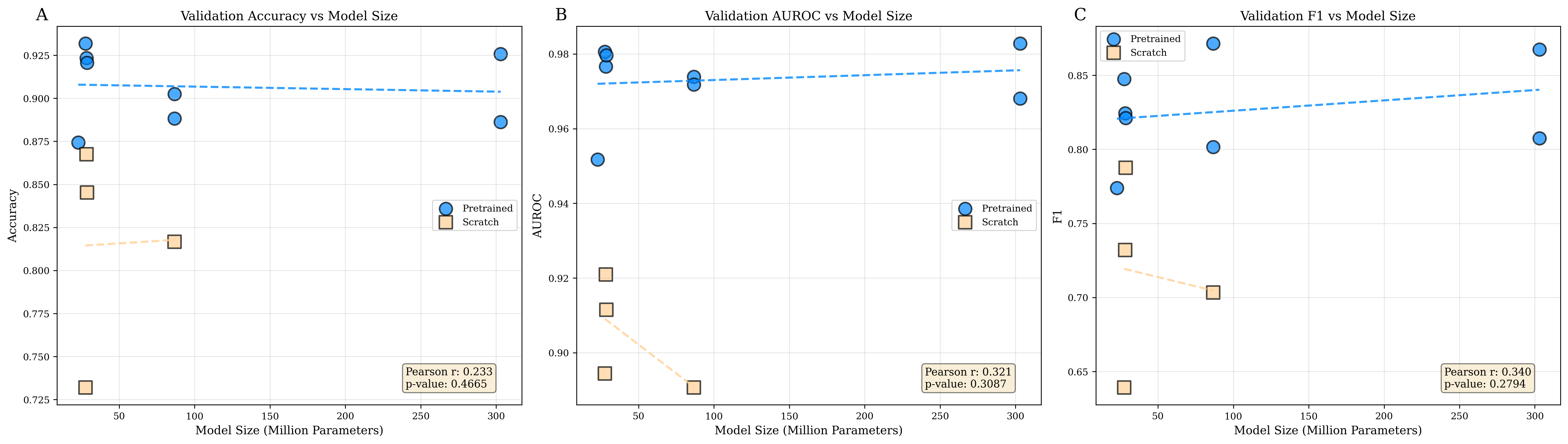}
\caption{GL: Multi-panel scatter plots for glaucoma detection showing 9.13 percentage point pretraining advantage. Pretrained (blue), scratch (green). Pearson correlation statistics (r, p-value) quantify the linear relationship between model size and performance. \textbf{(A) Accuracy vs Parameter Count}: Pretrained cluster tightly 0.87--0.93, scratch spread 0.73--0.86, with vertical separation intermediate between OCT (5.18\%) and DR (18.41\%). Performance saturates at 28M parameters (SwinV2-tiny, 0.9319); the 303M RETFound (0.9257) actually underperforms compact SwinV2-tiny, demonstrating that increased size provides no benefit for this task. Large AIROGS dataset (114,381 images) enables scratch models to reach respectable performance (up to 0.8675), yet all still underperform pretrained models, indicating ImageNet features provide universal benefits regardless of data abundance. No significant correlation between size and accuracy ($p = 0.47$) confirms that larger models do not systematically outperform smaller ones. \textbf{(B) AUROC vs Parameter Count}: Pretrained cluster 0.95--0.983, scratch 0.89--0.92. Similar saturation at 28M. No significant correlation ($p = 0.31$). \textbf{(C) F1-Score vs Parameter Count}: Pretrained 0.77--0.872, scratch 0.63--0.79. Wider spread reflects class imbalance in 3-class labels. No significant correlation ($p = 0.28$). Consistent message: compact pretrained models (27--29M) achieve optimal performance; 303M specialized models provide no advantage for this moderate-difficulty CFP task with clear visual features.}
\label{fig:performance_size_gl}
\end{figure*}

\begin{figure*}[!ht]
\centering
\includegraphics[width=0.85\textwidth]{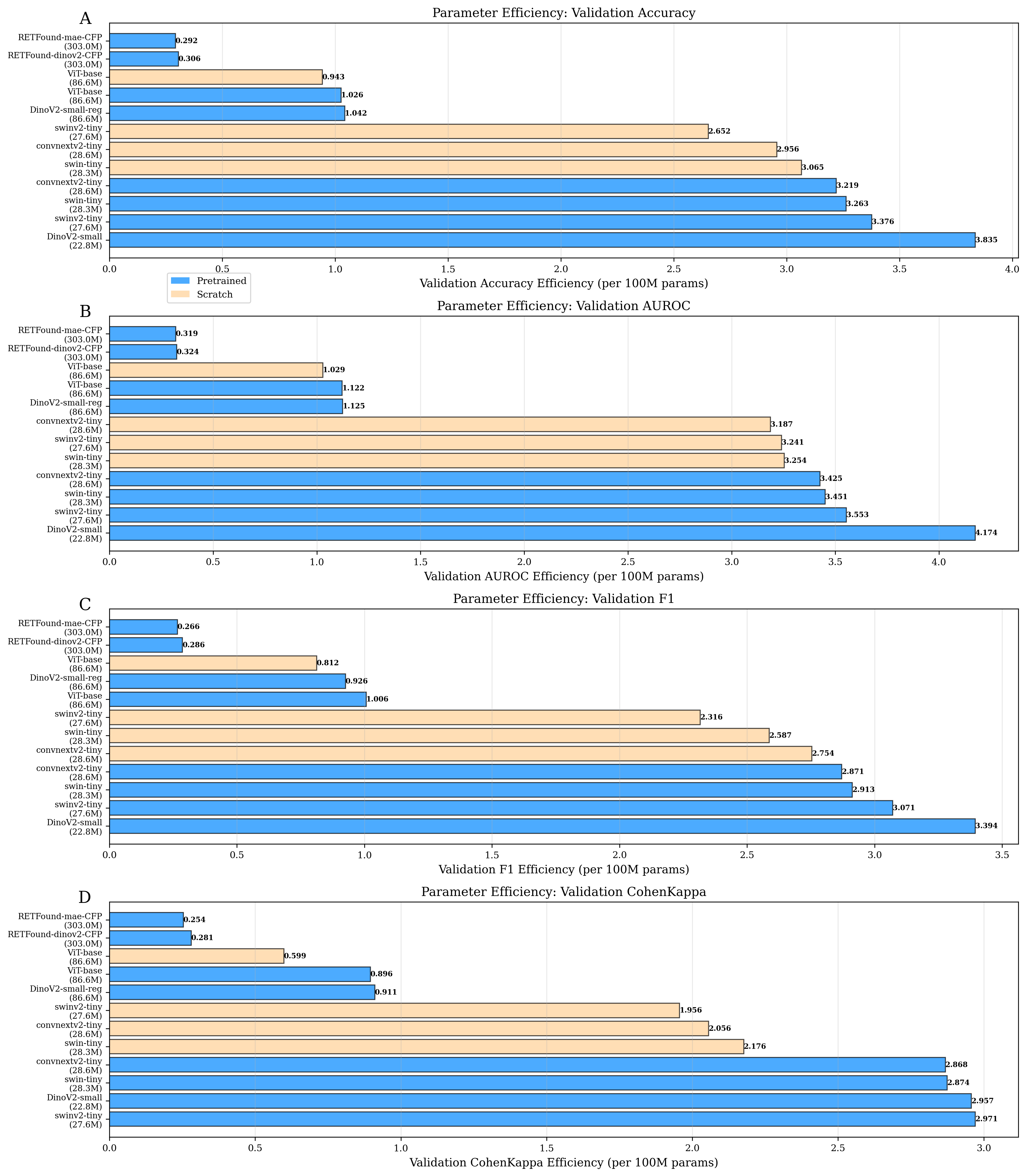}
\caption{GL: Multi-panel horizontal bar chart showing parameter efficiency for glaucoma detection. Y-axis: models with sizes; X-axis: efficiency [0.0--5.0]. \textbf{(A) Accuracy Efficiency}: DinoV2-small achieves 3.835 (87.43\%, 22.8M), offering strong performance for resource-constrained deployment with only 5.76 percentage points below top model. SwinV2-tiny provides optimal balance: 3.376 efficiency with 93.19\% accuracy, highest absolute performance for GL. Large domain-specific RETFound-DinoV2-CFP shows poor efficiency (0.306, 92.57\%, 303M), 12.4$\times$ lower than DinoV2-small while providing only 5.14 percentage points higher accuracy. Scratch models <3.065 due to low accuracy. \textbf{(B) AUROC Efficiency}: DinoV2-small leads (4.174), RETFound <0.324. \textbf{(C) F1 Efficiency}: Pretrained <3.394, scratch <2.754, RETFound <0.286. \textbf{(D) Kappa Efficiency}: DinoV2-small (2.957), pretrained <2.971, RETFound <0.281. Efficiency-accuracy trade-off suggests: for GL screening with limited resources, compact DinoV2-small offers excellent value; for maximum accuracy clinical settings, SwinV2-tiny provides near-optimal performance without 300M+ parameter computational burden.}
\label{fig:efficiency_gl}
\end{figure*}

\begin{figure*}[!ht]
\centering
\includegraphics[width=1.0\textwidth]{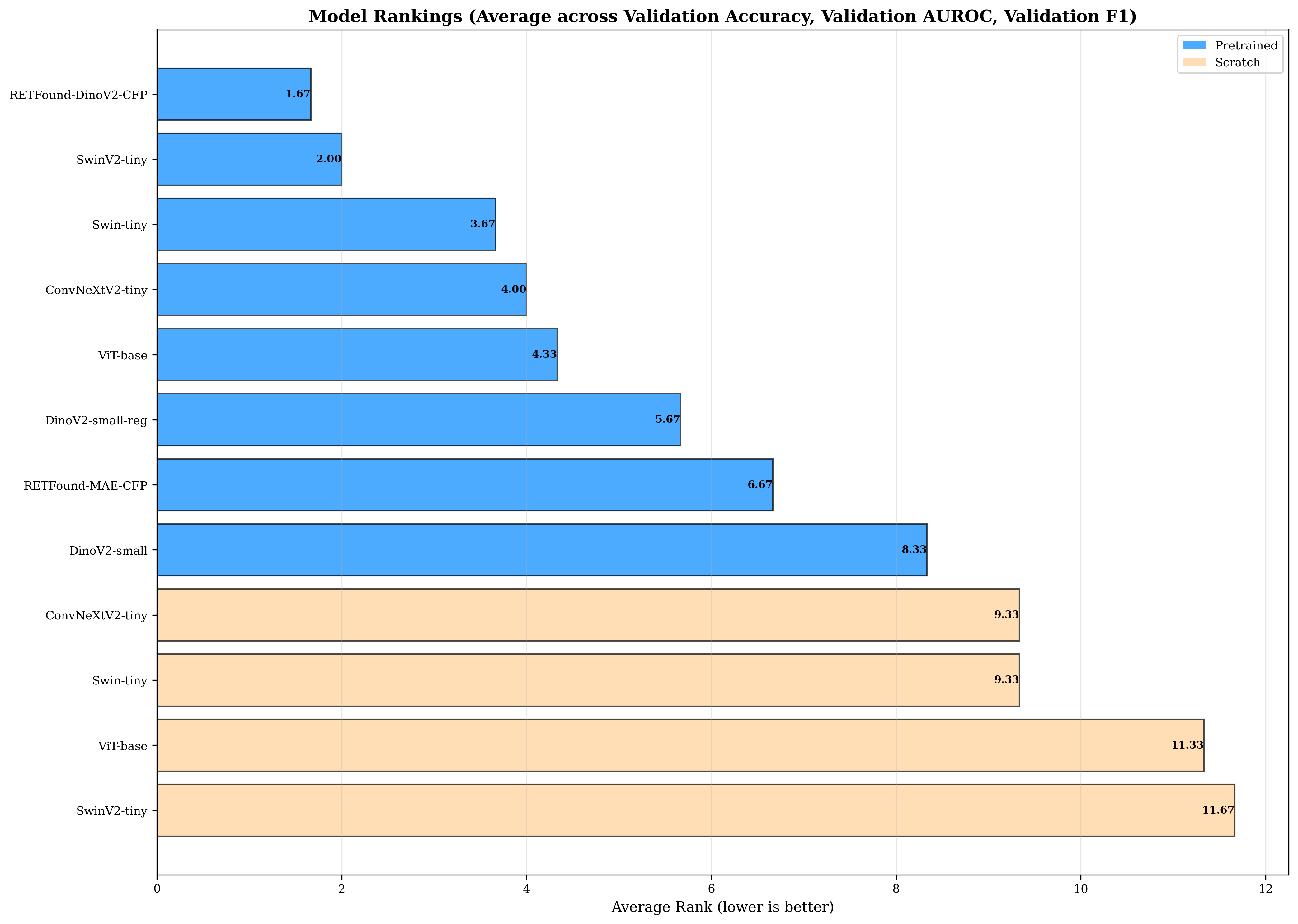}
\caption{GL: Aggregate model rankings for glaucoma detection. RETFound-DinoV2-CFP and SwinV2-tiny achieve nearly identical top average ranks (1.67 vs 2.00), but SwinV2-tiny achieves higher absolute accuracy (93.19\% vs 92.57\%) at 11$\times$ smaller size. The ranking stratification remains absolute: all eight pretrained models occupy ranks 1--8, while all four scratch-trained models occupy ranks 9--12, maintaining the consistent pattern observed. Unlike DR where RETFound ranked first, the GL rankings show that compact general-purpose models match specialized domain-specific models, indicating that retina-specific pretraining provides no systematic advantage for this 3-class glaucoma detection task. The compact DinoV2-small ranks 6th among pretrained models (similar to DME and OCT), demonstrating adequate but not optimal performance for moderate-difficulty CFP tasks.}
\label{fig:rankings_gl}
\end{figure*}

RETFound-DinoV2-CFP and SwinV2-tiny achieved nearly identical top performance (92.57\% vs.\ 93.19\%), with SwinV2 delivering slightly higher accuracy despite 11$\times$ fewer parameters. All pretrained models (ranks 1--8) outperformed scratch-trained models (ranks 9--12), maintaining the consistent pattern observed across all tasks (Figure~\ref{fig:rankings_gl}). Parameter efficiency analysis identified DinoV2-small (87.43\%, 22.8M, efficiency 3.83) as optimal for deployment (Figure~\ref{fig:efficiency_gl}). Pareto frontier analysis highlighted two models: dinov2-small (22.8M) and SwinV2-tiny (27.6M, 93.19\%), as shown in Figure~\ref{fig:pareto_gl}. Figure~\ref{fig:performance_size_gl} demonstrates the performance-size trade-offs across all models. Pearson correlation analysis revealed no significant linear relationship between model size and performance (accuracy: $p = 0.47$; AUROC: $p = 0.31$; F1-score: $p = 0.28$), further confirming that larger models do not provide systematic advantages for glaucoma detection.

\subsubsection{Architecture-Specific Observations}
Hierarchical Swin Transformers achieved the highest absolute performance, with SwinV2-tiny ranking first in accuracy (93.19\%) and Swin-tiny placing third (92.34\%). Unlike DR, domain-specific RETFound models did not provide clear advantages over ImageNet-pretrained architectures for glaucoma detection, suggesting the optic disc-centered AIROGS/PAPILA imaging may not require specialized retinal pretraining. The pretraining benefit varied substantially: SwinV2 improved by 19.9 percentage points from scratch, while ConvNeXt showed 7.54 percentage points improvement, suggesting hierarchical attention mechanisms benefit more from ImageNet initialization. See Appendix~\ref{app:gl_figures} for additional visualizations.

\subsection{Cross-Dataset Comparison}
\label{sec:results:comparison}

To understand how backbone initialization strategies and architectures generalize across different retinal imaging tasks, we compare results across all four datasets: OCT (8-class balanced), DME (3-class CFP), DR (5-class ordinal CFP with severe imbalance), and GL (3-class CFP). This analysis reveals task-dependent patterns in pretraining benefits, architecture robustness, and optimal model selection strategies.

\subsubsection{Pretraining Benefit Across Datasets}

The advantage of pretrained initialization varied substantially across tasks, correlating strongly with task difficulty and imaging modality (Table~\ref{tab:pretraining_comparison_all}). Pretraining benefits ranged from 5.18 percentage points (OCT) to 18.41 percentage points (DR), with the improvement magnitude inversely related to absolute task performance.

\begin{table*}[t]
\centering
\caption{Pretraining benefit across all four datasets, measured by difference in best validation accuracy. Pretraining advantage increases with task difficulty, from 5.18\% for balanced OCT classification to 18.41\% for challenging ordinal DR grading. All differences are statistically significant (p < 0.05).}
\label{tab:pretraining_comparison_all}
\small
\begin{tabular}{@{}lccccc@{}}
\toprule
\textbf{Dataset} & \textbf{Pretrained} & \textbf{Scratch} & \textbf{Difference} & \textbf{Best Model Acc} & \textbf{Modality} \\
\midrule
OCT (8-class, balanced) & 0.9694 $\pm$ 0.0088 & 0.9176 $\pm$ 0.0337 & +5.18\% & 97.96\% & OCT \\
DME (3-class) & 0.9813 $\pm$ 0.0209 & 0.8743 $\pm$ 0.0772 & +10.70\% & 99.24\% & CFP \\
GL (3-class) & 0.9066 $\pm$ 0.0217 & 0.8154 $\pm$ 0.0593 & +9.13\% & 93.19\% & CFP \\
DR (5-class ordinal) & 0.6713 $\pm$ 0.0289 & 0.4872 $\pm$ 0.0807 & +18.41\% & 71.15\% & CFP \\
\bottomrule
\end{tabular}
\end{table*}

Three key patterns emerged: (1) \textbf{Imaging modality effects}: CFP-based tasks (DME, DR, GL) showed larger pretraining benefits (9.13--18.41\%) compared to OCT (5.18\%), suggesting that natural image features from ImageNet transfer more effectively to color fundus photography than to grayscale cross-sectional OCT imaging. (2) \textbf{Task difficulty scaling}: The pretraining advantage increased with task complexity. DR (5-class ordinal with 8.22\% severe class) showed the largest benefit (18.41\%), while the balanced 8-class OCT task showed the smallest (5.18\%). (3) \textbf{Variance reduction}: Beyond accuracy improvements, pretraining substantially reduced performance variance across models, with scratch-trained models showing 2--4$\times$ higher standard deviations.

\subsubsection{Architecture Performance Consistency}

We assessed architecture robustness by evaluating performance consistency across all four tasks (Table~\ref{tab:architecture_consistency}). Microsoft's Swin Transformer variants demonstrated the most consistent high performance, appearing in the top-3 models for all four datasets. Specifically, Swin-tiny achieved top-2 rankings on three tasks (OCT: rank 1, DME: rank 1, GL: rank 2) and strong performance on DR (rank 2).

\begin{table*}[t]
\centering
\caption{Architecture family performance consistency across all datasets. Rankings based on best validation accuracy within each task. Hierarchical architectures (Swin, ConvNeXt) show the most consistent performance, while domain-specific RETFound models excel only on the most challenging task (DR).}
\label{tab:architecture_consistency}
\small
\begin{tabular}{@{}lccccl@{}}
\toprule
\textbf{Architecture} & \textbf{OCT Rank} & \textbf{DME Rank} & \textbf{DR Rank} & \textbf{GL Rank} & \textbf{Mean Rank} \\
\midrule
SwinV2-tiny (27.6M) & 1 & 1 & 2 & 2 & 1.50 \\
RETFound-DinoV2-CFP (303M) & 6 & 2 & 1 & 1 & 2.50 \\
ConvNeXtV2-tiny (28.6M) & 2 & 5 & 5 & 4 & 4.00 \\
Swin-tiny (28.3M) & 3 & 8 & 3 & 3 & 4.25 \\
DinoV2-small-reg (86.6M) & 5 & 4 & 4 & 6 & 4.75 \\
ViT-base (86.6M) & 7 & 7 & 7 & 5 & 6.50 \\
\bottomrule
\end{tabular}
\end{table*}

ConvNeXt variants showed strong but less consistent performance (mean rank 4.00), excelling on OCT (rank 2) but showing moderate performance on CFP tasks. Domain-specific RETFound models exhibited task-dependent behavior: they achieved top rankings on the most challenging tasks (DR: rank 1, GL: rank 1, DME: rank 2) but underperformed on the easier, balanced OCT classification (rank 6), suggesting that retina-specific pretraining likely provides advantages primarily when fine-grained disease discrimination is required.

Notably, the compact DinoV2-small (22.8M parameters) maintained competitive performance across all tasks despite being 3--13$\times$ smaller than top models. It achieved parameter efficiency values of 4.20 (OCT), 4.08 (DME), 2.71 (DR), and 3.83 (GL) accuracy per 100M parameters, demonstrating that smaller models can deliver strong performance relative to their size.

\FloatBarrier
\clearpage
\section{Discussion}
\label{sec:discussion}

To our knowledge, this study provides the first systematic evaluation of backbone initialization strategies and architectures for retinal disease classification across four distinct imaging tasks, and two imaging modalities. Our comprehensive benchmarking reveals task-dependent patterns in the value of pretraining, architecture selection strategies, and the role of domain-specific initialization, challenging prevailing assumptions in foundation model development.

\subsection{Key Findings and Their Implications}

\subsubsection{Pretraining Provides Consistent but Task-Dependent Benefits}

Pretrained initialization significantly outperformed scratch training across all four datasets ($p < 0.05$), with accuracy improvements ranging from 5.18\% (OCT) to 18.41\% (DR). Two critical patterns emerged: First, the pretraining benefit scaled inversely with task performance. Easier tasks with clearer visual patterns (DME on CFP: 99.24\% best accuracy) showed moderate pretraining advantages (10.70\%), while challenging tasks with subtle inter-class boundaries (DR on CFP: 71.15\% best accuracy) demonstrated the largest benefits (18.41\%). This suggests that pretrained features are especially valuable when limited task-specific data must support fine-grained discrimination.

Second, imaging modality substantially influenced transfer learning effectiveness. Color fundus photography classification tasks (DME, DR, GL) showed larger pretraining benefits (9.13--18.41\%) than OCT imaging (5.18\%), indicating that ImageNet's natural image features likely align more closely with RGB fundus photographs than grayscale cross-sectional OCT scans. This reveals that imaging modality characteristics determine the necessity of large-scale pretraining infrastructure, contradicting assumptions that all medical imaging domains uniformly require extensive pretrained foundation models.

\subsubsection{Compact Models Dominate Pareto Frontiers}

Across all tasks, Pareto-optimal models clustered in the 22.8--29M parameter range, with larger models (86.6M, 303M) failing to provide proportional accuracy gains. The SwinV2-tiny (27.6M) achieved top-1 rankings on three of four tasks, while the DinoV2-small (22.8M) delivered 93--97\% of best-model performance at 3--13$\times$ smaller size. This directly contradicts the "bigger is better" paradigm driving current foundation model development, demonstrating that compact hierarchical architectures achieve near-optimal performance without requiring the computational infrastructure investments associated with 300M+ parameter models.

The dominance of smaller models likely reflects the relatively constrained visual vocabulary of retinal imaging compared to natural images. Unlike ImageNet's 1000 diverse object categories, retinal disease classification focuses on specific pathological features such as hemorrhages, exudates, drusen, layer disruptions, that may not require the representational capacity of 300M-parameter models.

\subsubsection{Domain-Specific Pretraining Shows Task-Selective Value}

Domain-specific RETFound models (303M parameters, pretrained on retinal images) demonstrated task-dependent advantages. They achieved top performance only on DR (71.15\%, rank 1), the most challenging task with extreme class imbalance and ordinal severity grading. On all other tasks, smaller ImageNet-pretrained models matched or exceeded RETFound performance (OCT: ConvNeXtV2-tiny 97.96\% vs RETFound-MAE-OCT 97.29\%; DME: SwinV2-tiny 99.24\% vs RETFound 99.06\%; GL: SwinV2-tiny 93.19\% vs RETFound-DinoV2-CFP 92.57\%).

This contradicts the research assumption that domain-specific foundation models provide universal advantages for medical imaging. The substantial computational investment required for large-scale self-supervised retinal pretraining justifies itself potentially only for the most challenging task (DR severity grading), while general-purpose ImageNet pretraining delivers equivalent performance on three of four tasks.

\subsubsection{Hierarchical Architectures Provide Robust Cross-Task Performance}

Swin Transformers demonstrated superior consistency across imaging modalities and task difficulties, with SwinV2-tiny achieving mean rank 1.50 across all datasets. This robustness likely stems from their hierarchical, multi-scale feature extraction, capturing both local pathological details such as microaneurysms, small drusen, and global structural patterns such as optic disc morphology, vessel organization, critical for comprehensive retinal assessment.

In contrast, standard Vision Transformers (ViT) showed inconsistent performance and the largest pretraining dependency (OCT: 96.86\% pretrained vs 88.04\% scratch), suggesting their global self-attention mechanism struggles without initialization from large-scale pretraining. These findings indicate that architectural inductive biases matter more than model scale, with 27M-parameter hierarchical transformers outperforming 86M+ parameter standard ViTs across diverse task characteristics.

\subsubsection{Task-Specific Insights}

Optimal model selection strategies varied systematically by imaging modality and task characteristics:

\textbf{OCT (Balanced Multi-Class):} Compact hierarchical architectures (Swin Transformers, ConvNeXt) in the 27--29M parameter range dominated the Pareto frontier, achieving 97.64--97.96\% accuracy. Domain-specific RETFound models (303M parameters) provided no advantage over ImageNet-pretrained models, suggesting that the structural patterns in OCT imaging (layer segmentation, fluid accumulation) are adequately captured by general-purpose visual features. The balanced class distribution (2,300 samples/class) enabled effective learning even from scratch (91.76\% mean accuracy), resulting in the smallest pretraining benefit observed.

\textbf{DME (Few-Class CFP):} This task achieved the highest absolute performance (99.24\% best accuracy), with SwinV2-tiny outperforming all competitors including larger domain-specific models. The combination of (1) simplified 3-class grading, (2) clear visual markers (hard exudate location), and (3) effective ImageNet feature transfer made this the most tractable task. Scratch-trained models struggled dramatically (87.43\% mean accuracy), showing the largest performance variance (std: 0.0772) and highlighting the critical importance of pretraining for CFP-based classification with limited training data (2,264 images).

\textbf{DR (Ordinal CFP with Severe Imbalance):} This task uniquely favored domain-specific pretraining, with RETFound-DinoV2-CFP (71.15\%) outperforming all ImageNet-pretrained models. The extreme class imbalance (Class 3: 8.22\%), subtle inter-class boundaries (ordinal severity grading), and need to detect fine-grained microaneurysms/hemorrhages likely benefit from retina-specific feature representations learned during self-supervised pretraining on large-scale fundus datasets. The largest pretraining benefit observed (18.41\%) and lowest absolute performance (71.15\%) indicate this as the most challenging task, where specialized inductive biases become valuable.

\textbf{GL (3-Class CFP):} Results fell between DME (easy) and DR (hard), with SwinV2-tiny (93.19\%) marginally outperforming RETFound models. The large dataset size (114,381 images from AIROGS) and relatively clear optic disc-centered imaging enabled strong performance from ImageNet-pretrained hierarchical architectures. Unlike DR, the simpler 3-class grading structure and abundant training data reduced the need for domain-specific pretraining, with general-purpose Swin Transformers providing the optimal accuracy-size trade-off.

\subsection{Implications for Foundation Model Development}

Our findings highlight cases where specialized, large-scale foundation models are necessary from those where compact, general-purpose models are sufficient:

\begin{enumerate}
  \item \textbf{Compact models challenge scaling assumptions}: Hierarchical transformers (27--29M parameters) like SwinV2-tiny achieve top-2 performance on three of four tasks, questioning whether the computational investment in 300M+ parameter foundation models yields proportional benefits for most retinal imaging applications.

  \item \textbf{Domain-specific pretraining shows limited advantage}: Specialized retinal foundation models (RETFound, 303M) outperform general-purpose ImageNet initialization only for the most challenging ordinal grading task (DR). For three of four tasks, ImageNet pretraining delivers equivalent or superior performance at 10$\times$ smaller model size, suggesting the drive toward domain-specific foundation models may be overestimated.

  \item \textbf{Task difficulty predicts specialization value}: The extreme class imbalance and fine-grained discrimination required for DR severity grading represents the boundary case where specialized pretraining may justify its cost. More tractable clinical tasks achieve near-optimal performance with compact general-purpose models.

  \item \textbf{Modality-specific resource allocation}: CFP-based applications benefit more from pretraining investments (9.13--18.41\% gains) than OCT applications (5.18\%), suggesting resource allocation strategies should account for imaging modality characteristics rather than assuming uniform foundation model requirements.
\end{enumerate}

\subsection{Limitations and Future Directions}

Several limitations warrant consideration. First, our evaluation used single institutional/competition datasets for each task, limiting generalizability assessment. External validation on datasets from different imaging devices, geographic populations, and acquisition protocols would strengthen conclusions about architecture and initialization strategy robustness. Second, we evaluated models based on classification accuracy; complementary analyses such as calibration, uncertainty quantification, and failure mode characterization represent valuable directions for future work.

Third, we evaluated only cross-entropy loss with standard training procedures. Recent advances in ordinal regression losses, focal loss variants for extreme imbalance, and contrastive learning approaches may further improve performance, especially for challenging tasks like DR grading. Fourth, our analysis considered only final model selection; convergence speed, sample efficiency, and performance under data scarcity (few-shot scenarios) represent important practical considerations not addressed here.

Future work could investigate: (1) multi-task learning across disease types to leverage shared retinal feature representations, (2) hybrid architectures combining hierarchical transformers with domain-specific inductive biases (e.g., vessel-aware attention), (3) systematic analysis of pretraining dataset composition to identify minimal sufficient pretraining for retinal imaging, and (4) development of task-difficulty metrics that predict when domain-specific vs general-purpose pretraining is warranted.

Finally, our focus on model architecture and initialization isolates these factors from confounding variables. Nonetheless, comprehensive system evaluation requires additional consideration of data curation quality, annotation consistency, and algorithmic fairness across demographic subgroups, aspects that extend beyond this architectural benchmarking study.

\newpage
\section{Conclusion}
\label{sec:conclusion}

This study provides the first comprehensive benchmarking of backbone initialization strategies and architectures across multiple retinal imaging modalities and disease classification tasks. Through systematic evaluation of 13 model configurations on four distinct datasets (OCT, DME, DR, GL), we challenge prevailing research assumptions about foundation model development for medical imaging.

Our key findings demonstrate that pretrained initialization provides universal benefits, with accuracy improvements ranging from 5.18\% to 18.41\% depending on task difficulty and imaging modality. Critically, the pretraining advantage scales with task complexity. Challenging ordinal classification with class imbalance (DR) benefits most from pretrained features, while balanced multi-class tasks (OCT) show smaller but still significant improvements. Color fundus photography tasks showed consistently larger pretraining benefits than OCT imaging, indicating that ImageNet's natural image statistics transfer more effectively to RGB fundus photographs.

Our hypothesis was confirmed: contrary to prevailing trends toward ever-larger models, we found that compact hierarchical architectures (27--29M parameters) dominated Pareto frontiers across all tasks. Microsoft's SwinV2-tiny achieved top-2 performance on all four datasets, while the 22.8M-parameter DINOv2-small delivered 93--97\% of best-model accuracy at 3--13$\times$ smaller size. This fundamentally challenges the research assumption that larger foundation models provide proportional performance gains, demonstrating that compact architectures achieve near-optimal results without massive computational infrastructure investments.

Domain-specific retinal pretraining showed task-selective value, achieving top performance only on the most challenging task (DR severity grading) while providing no advantages over ImageNet pretraining for more tractable problems. This suggests that the substantial computational investment required for large-scale self-supervised retinal pretraining is justified primarily when fine-grained pathology discrimination is critical, not as a universal strategy.

Our findings contradict three prevailing research assumptions: (1) Large foundation models (300M+ parameters) are necessary for state-of-the-art medical imaging performance; nonetheless our results show compact hierarchical transformers (27--29M parameters) achieve top-1 rankings on three of four tasks. (2) Domain-specific pretraining universally outperforms general-purpose ImageNet initialization; however we demonstrate ImageNet pretraining suffices for three of four tasks, with specialized models justified only for the most challenging ordinal grading scenario. (3) Bigger models consistently outperform smaller architectures; and yet Pareto frontier analysis reveals 22.8--29M parameter models dominate efficiency-accuracy trade-offs across all datasets.

These findings suggest the field needs to reconsider current trajectories in foundation model development. The task-dependent patterns we identify (when compact models suffice, when domain-specific pretraining justifies its cost, and which architectural inductive biases transfer effectively) provide an evidence base for resource allocation decisions in medical imaging AI. As the field advances toward multi-task learning and hybrid architectures, systematic benchmarking across task characteristics will be essential for distinguishing genuine architectural advances from incremental scaling of computational resources.

\newpage
\section*{Author Contributions}
D.I. conceived the study, designed experiments, wrote the manuscript, performed data analysis, and created visualizations. T.S. and G.M.S. provided clinical expertise and reviewed the manuscript. R.S. conceived the study, reviewed the manuscript, and submitted the manuscript. All authors reviewed and approved the final manuscript.

\section*{Ethics and Consent to Participate}
This study exclusively utilized publicly available, de-identified datasets. No ethics approval was required as all data were previously collected with appropriate institutional approvals by the original dataset creators, and our analysis involved only computational modeling on de-identified images without any direct patient contact or identifiable information.

\section*{Funding}
This work was supported by the Werner H. Spross-Stiftung. There is no grant number associated with this funding.

\section*{Acknowledgements}
We acknowledge the Spross Research Institute for providing computational resources that enabled this study.

\section*{Competing Interests}
D.I., T.S., and R.S. declare no competing interests. G.M.S. serves as a consultant for Apellis, Allergan, Bayer, Carl Zeiss, and Roche.

\FloatBarrier
\twocolumn
\bibliographystyle{unsrt}
\bibliography{refs}

\onecolumn
\appendix

\section{Supplementary Material}
\label{app:figures}

This appendix provides supplementary materials supporting the paper, including detailed methodology, reproducibility documentation, and comprehensive visualizations. Section~\ref{app:extended_methods} presents extended materials and methods with full experimental details. Subsequent sections provide the HuggingFace model identifier mapping and detailed visualization figures for all four retinal disease classification tasks.

\twocolumn
\subsection{Extended Materials and Methods}
\label{app:extended_methods}

This section provides comprehensive experimental details expanding on the compact Methods section in the body of the paper.

\subsubsection{Dataset Details}
\label{app:dataset_details}

\textbf{OCT}: The Retinal OCT Image Classification - C8 dataset provides 24,000 images across 8 disease categories with perfect class balance (2,300 training, 350 validation, 350 test images per class). We used the provided splits without modification \cite{he2023interpretable}.

\textbf{Diabetic Retinopathy (DR)}: We used the EAM unified diabetic retinopathy dataset, which combines EyePACS, APTOS, APTOS (Gaussian Filtered), and Messidor datasets with manual data augmentation (increasing dataset size by approximately 55\%) and standardized resizing to 600x600 pixels. The dataset applies the International Clinical Diabetic Retinopathy (ICDR) severity scale for 5-class grading (0: No DR, 1: Mild NPDR, 2: Moderate NPDR, 3: Severe NPDR, 4: Proliferative DR). The dataset exhibits significant class imbalance with Class 3 (Severe) representing 8.22\% of images. We used the provided stratified 80/10/10 split preserving ordinal severity relationships \cite{canipek2025eyepacs}.

\textbf{Diabetic Macular Edema (DME)}: We combined IDRiD (516 images) and Messidor-2 (1,748 images) with compatible 3-class severity labels (Grade 0: No DME, Grade 1: Mild DME with distant hard exudates, Grade 2: Severe DME with hard exudates near macula). Stratified 70/15/15 splitting ensured balanced representation across clinical sources \cite{porwal2018indian,decenciere2014feedback}.

\textbf{Glaucoma (GL)}: We combined AIROGS (113,893 population screening images) and PAPILA (488 bilateral diagnostic images) into a unified 3-class dataset. AIROGS binary labels (0: non-referable, 1: referable glaucoma) and PAPILA 3-class labels (0: non-glaucomatous, 1: glaucomatous, 2: suspect) were mapped to a shared label space \{0, 1, 2\}. Stratified 70/15/15 splitting maintained representation from both screening and diagnostic contexts \cite{devente23airogs,kovalyk2022papila}.

\begin{table*}[h]
\centering
\caption{Overview of retinal imaging datasets used in this study. \textbf{OCT}: Optical Coherence Tomography retinal imaging. \textbf{DR}: Diabetic Retinopathy severity grading (ICDR scale 0--4). \textbf{DME}: Diabetic Macular Edema severity grading (0--2). \textbf{GL}: Glaucoma detection combining screening and diagnostic contexts.}
\label{tab:datasets}
\small
\begin{tabular}{@{}lp{2.5cm}cccp{5.5cm}@{}}
\toprule
\textbf{Task} & \textbf{Source} & \textbf{Images} & \textbf{Classes} & \textbf{Split} & \textbf{Key Characteristics} \\
\midrule
\textbf{OCT} & Retinal OCT C8 \cite{he2023interpretable} & 24,000 & 8 & Provided & Balanced: 8 disease categories (AMD, CNV, CSR, DME, DR, Drusen, MH, Normal); 2,300/350/350 per class \\
\midrule
\textbf{DR} & EAM \cite{canipek2025eyepacs}: Eyepacs, Aptos, Messidor & 143,669 & 5 & 80/10/10 & Imbalanced ICDR scale (0--4): No DR 47.87\%, Moderate 16.03\%, Mild 21\%, Proliferative 6.9\%, Severe 8.22\% \\
\midrule
\textbf{DME} & IDRiD \cite{porwal2018indian} + Messidor-2 \cite{decenciere2014feedback} & 2,264 & 3 & 70/15/15 & Combined sources; Grades 0--2 based on hard exudate distance from macula \\
\midrule
\textbf{GL} & AIROGS \cite{devente23airogs} + PAPILA \cite{kovalyk2022papila} & 114,381 & 3 & 70/15/15 & Combined dataset: AIROGS binary (0: non-referable, 1: referable) + PAPILA 3-class (0: non-glaucomatous, 1: glaucomatous, 2: suspect); unified to labels \{0, 1, 2\} \\
\bottomrule
\end{tabular}
\end{table*}

\subsubsection{Training Infrastructure and Procedures}
\label{app:training_details}

To ensure reproducibility and fair comparison across all models and datasets, we developed a unified training platform implementing consistent training procedures, optimization configurations, and evaluation protocols.

\textbf{Optimization Configuration}: All models were trained using the AdamW optimizer \cite{loshchilov2017decoupled} with cosine learning rate schedules \cite{loshchilov2016sgdr} including warmup phases. Learning rates were tuned per-model and initialization strategy based on preliminary experiments (Table~\ref{tab:training_hyperparams}). Weight decay was set to 0.05 for all configurations, with proper handling to exclude bias and LayerNorm parameters from regularization, which is critical for transformer model stability \cite{zhang2019lookahead}.

\begin{table}[h]
\centering
\caption{Model-specific learning rates for pretrained and from-scratch training. Following established practices \cite{steiner2022trainvitdataaugmentation,kumar2023finetunevisionmodelssgd}, pretrained models use lower learning rates to preserve learned representations, while from-scratch models tolerate higher rates. All rates within standard ranges for AdamW optimization of vision transformers.}
\label{tab:training_hyperparams}
\small
\begin{tabular}{@{}lcc@{}}
\toprule
\textbf{Model} & \textbf{Pretrained} & \textbf{Scratch} \\
\midrule
ViT-base & 3e-4 & 5e-4 \\
DINOv2-small & 3e-4 & N/A \\
DINOv2-with-registers & 2e-4 & N/A \\
Swin-tiny & 5e-4 & 5e-4 \\
SwinV2-tiny & 4e-4 & 5e-4 \\
ConvNeXtV2-tiny & 1e-3 & 1e-3 \\
RETFound (all variants) & 3e-4 & N/A \\
\bottomrule
\end{tabular}
\end{table}

\textbf{Batch Size and Gradient Accumulation}: To maintain computational fairness across models with different memory requirements, we employed gradient accumulation to achieve a consistent effective batch size of 256 across all experiments. Physical batch sizes were set to 64 for most models, with 48 for models using 256$\times$256 input resolution (SwinV2-tiny). Gradient accumulation steps were computed as $\lceil 256 / \text{batch\_size} \rceil$, ensuring identical optimization dynamics regardless of GPU memory constraints.

\textbf{Training Stability Measures}: Three mechanisms ensured stable training: (1) Mixed-precision training \cite{micikevicius2017mixed} with bfloat16 autocasting reduced memory usage while maintaining numerical stability compared to float16. (2) Gradient clipping \cite{pascanu2013difficulty} with maximum norm 1.0 prevented exploding gradients, especially critical for scratch-trained transformers. (3) Learning rate warmup over 10\% of total training steps enabled smooth optimization initialization, preventing early training instability observed in preliminary experiments without warmup \cite{goyal2018accuratelargeminibatchsgd,vaswani2023attentionneed,kalra2024warmuplearningrateunderlying}.

\textbf{Training Duration and Checkpointing}: All models trained for up to 100 epochs. We monitored validation accuracy and saved checkpoints for: (1) the epoch achieving best validation accuracy, and (2) optionally per-epoch weights for convergence analysis. No early stopping was employed; final results report the best validation accuracy achieved across all 100 epochs, ensuring fair comparison without hyperparameter-dependent stopping criteria.

\textbf{Data Augmentation}: To preserve comparability and avoid confounding augmentation effects with architecture/initialization differences, we applied minimal augmentation: bilinear resize and center crop to model-specific input resolutions (224$\times$224 or 256$\times$256), followed by normalization using each model's ImageNet statistics (mean/std extracted from Hugging Face AutoImageProcessor). No horizontal flips, rotations, or color jittering were applied to training data, maintaining consistency across all experiments. This conservative augmentation strategy isolates the effects of model architecture and pretraining from data augmentation benefits.

\textbf{Reproducibility}: Deterministic training was enforced through: (1) fixed random seed (42) for Python, NumPy, and PyTorch RNGs at experiment start, (2) deterministic cuDNN convolution algorithms when available, and (3) consistent data loading with fixed worker seeds. All experiments used identical code paths from a unified training platform, with per-experiment configurations specified declaratively to prevent implementation variations.

\textbf{Loss Function}: Cross-entropy loss \cite{bishop2006pattern} was used across all experiments without class weighting or focal loss modifications. This provides a consistent optimization objective, with performance differences attributable solely to architecture and initialization rather than task-specific loss engineering.

\textbf{Computational Setup}: All experiments were conducted on NVIDIA GPUs with CUDA support. Batch processing leveraged asynchronous data loading with 8 workers, prefetch factor of 4, and persistent workers to maximize GPU utilization. cuDNN benchmark mode was enabled to auto-select optimal convolution algorithms per model. Training times ranged from 2--8 hours per 100-epoch run depending on model size and dataset.

\subsubsection{Evaluation Metrics Details}
\label{app:metrics_details}

\textbf{Accuracy (Acc)}: Top-1 classification accuracy, computed as the proportion of correctly classified samples. Range: [0, 1], higher is better.

\textbf{AUROC (macro)}: Area Under the Receiver Operating Characteristic curve \cite{hanley1982meaning,fawcett2006introduction}, macro-averaged across classes. This threshold-independent metric measures the ability to distinguish between classes, with equal weight given to each class regardless of prevalence. Range: [0, 1], where 0.5 indicates random chance and 1.0 indicates perfect discrimination. Higher is better.

\textbf{F1-score (macro)}: Harmonic mean of precision and recall ($F1 = 2 \cdot \frac{\text{precision} \cdot \text{recall}}{\text{precision} + \text{recall}}$) \cite{powers2011evaluation}, macro-averaged across classes. Balances false positives and false negatives, with equal weight per class. Range: [0, 1], higher is better.

\textbf{Cohen's Kappa ($\kappa$)}: Agreement measure adjusted for chance agreement \cite{cohen1960coefficient}, defined as $\kappa = \frac{p_o - p_e}{1 - p_e}$ where $p_o$ is observed agreement and $p_e$ is expected agreement by chance. Accounts for class imbalance. Range: [-1, 1], where $\kappa = 0$ indicates chance-level performance, $\kappa = 1$ indicates perfect agreement, and negative values indicate worse than chance. Higher is better.

\textbf{Mean Average Precision (mAP)}: Average of precision values at each recall threshold, averaged across all classes. Summarizes the precision-recall curve. Range: [0, 1], higher is better.

\subsubsection{Statistical Analysis Details}
\label{app:statistical_details}

\textbf{Pretrained vs.\ Scratch Comparison}: Independent Mann-Whitney U tests \cite{mann1947test} were used to assess whether performance differences between pretrained and scratch-trained models were statistically significant. A statistical significance threshold of $\alpha = 0.05$ was used.

\textbf{Model Ranking}: Models were ranked separately on three key metrics (accuracy, AUROC, F1-score), with rank 1 assigned to the best-performing model for each metric. The average rank across the three metrics provides an overall performance indicator, reducing sensitivity to single-metric optimization.

\textbf{Parameter Efficiency}: Defined as validation accuracy per 100M parameters: $\text{Efficiency} = \frac{\text{Accuracy}}{\text{Parameters (in 100M)}} \times 100$. This metric identifies models that achieve high performance with minimal computational cost, important for resource-constrained deployment.

\textbf{Pareto Frontier Analysis}: Models were evaluated in the accuracy-size trade-off space \cite{miettinen1999nonlinear}. A model is Pareto-optimal (on the frontier) if no other model achieves higher accuracy without using more parameters. Models not on the frontier are strictly dominated and should not be selected under any parameter budget constraint.

\textbf{Pearson Correlation Analysis}: Pearson correlation coefficients ($r$) \cite{pearson1895notesonregressionand} were computed to assess the linear relationship between model parameter count and performance metrics (accuracy, AUROC, F1-score). The correlation coefficient ranges from -1 to +1, where $|r| = 1$ indicates perfect linear relationship, $r = 0$ indicates no linear correlation, and the sign indicates direction (positive: both variables increase together; negative: one increases as the other decreases). Statistical significance was tested using two-tailed tests at $\alpha = 0.05$. Non-significant correlations ($p > 0.05$) indicate that model size does not have a statistically significant linear relationship with performance, providing evidence against the assumption that larger models systematically outperform smaller architectures.

\FloatBarrier
\onecolumn
\subsection{HuggingFace Model Identifiers}
\begin{table*}[!ht]
\centering
\caption{Mapping between model names used in this study and their corresponding HuggingFace repository identifiers. All pretrained model weights were obtained from the HuggingFace model hub using these exact identifiers to ensure reproducibility. ImageNet-1k pretrained models use publicly available weights from their respective organizations (Google, Facebook/Meta, Microsoft). RETFound models use domain-specific weights pretrained on retinal imaging datasets using either Masked Autoencoder (MAE) or DINOv2 self-supervised learning approaches.}
\label{tab:backbones_hf}
\small
\begin{tabular}{@{}lll}
\toprule
\textbf{Model} & \textbf{HuggingFace Name} & \textbf{Pretraining Source} \\
\midrule
ViT-base & google/vit-base-patch16-224 & ImageNet-1k (supervised) \\
\midrule
DinoV2-small & facebook/dinov2-small-imagenet1k-1-layer & ImageNet-1k (DINOv2) \\
\midrule
DinoV2-small-reg & facebook/dinov2-with-registers-base & ImageNet-1k (DINOv2) \\
\midrule
Swin-tiny & microsoft/swin-tiny-patch4-window7-224 & ImageNet-1k \\
\midrule
SwinV2-tiny & microsoft/swinv2-tiny-patch4-window8-256 & ImageNet-1k \\
\midrule
ConvNeXtV2-tiny & facebook/convnextv2-tiny-1k-224 & ImageNet-1k \\
\midrule
RETFound-MAE-OCT & iszt/RETFound\_mae\_natureOCT & Retinal images (MAE) \\
\midrule
RETFound-DinoV2-CFP & iszt/RETFound\_dinov2\_meh & Retinal images (DINOv2) \\
\midrule
RETFound-MAE-CFP & iszt/RETFound\_mae\_meh & Retinal images (MAE) \\
\bottomrule
\end{tabular}
\end{table*}

\FloatBarrier
\newpage
\subsection{OCT Classification Figures}
\label{app:oct_figures}

\FloatBarrier
\subsection{DME Classification Figures}
\label{app:dme_figures}

\FloatBarrier
\subsection{DR Classification Figures}
\label{app:dr_figures}

\FloatBarrier
\subsection{GL Classification Figures}
\label{app:gl_figures}

\end{document}